\documentclass[fleqn,usenatbib,usedcolumn]{mnras}
\usepackage{mathptmx}

\usepackage[T1]{fontenc}
\usepackage{ae,aecompl}

\usepackage{graphicx}	
\usepackage{amsmath}	
\usepackage{amssymb}	

\newcommand{\um}{$\mu$m}

\newcommand{\ngc}{{NGC~1068}}
\newcommand{\C}{\textsc{Clumpy}}

\title[MIR imaging- and spectro-polarimetry of NGC 1068]{Mid-infrared imaging- and spectro-polarimetric subarcsecond observations of NGC 1068}

\author[Lopez-Rodriguez et al.]{
E. Lopez-Rodriguez$^{1,2}$\thanks{E-mail: \href{mailto:enrique.lopezrodriguez@utexas.edu}{enrique.lopezrodriguez@utexas.edu}},
C. Packham$^{3,4}$, 
P.~F.~Roche$^{5}$, 
A. Alonso-Herrero$^{3,5,6}$,
\newauthor 
T. D\'iaz-Santos$^{7,8}$,
R. Nikutta$^{9}$,
O. González-Martín$^{10}$,
C. A. \'Alvarez$^{11,12}$,
P. Esquej$^{13}$,
\newauthor
J.~M. Rodríguez Espinosa$^{11,12}$, 
E. Perlman$^{14}$,
C. Ramos-Almeida$^{11,12}$,    
C.~M. Telesco$^{15}$ 
\\
	$^{1}$Department of Astronomy, University of Texas at Austin, 1 University Station C1400, Austin, TX 78712, USA \\
	$^{2}$McDonald Observatory, University of Texas at Austin, Austin, TX 78712, USA \\
	$^{3}$Department of Physics \& Astronomy, University of Texas at San Antonio, One UTSA Circle, San Antonio, TX 78249, USA \\
	$^{4}$National Astronomical Observatory of Japan, Mitaka, Tokyo 181-8588, Japan \\
	$^{5}$Astrophysics, Department of Physics, University of Oxford, DWB, Keble Road, Oxford OX1 3RH, UK \\
	$^{6}$Instituto de F\'isica de Cantabria, CSIC-UC, E-39005 Santander, Spain \\
	$^{7}$Infrared Processing \& Analysis Center, MS 100-22, California Institute of Technology, Pasadena, CA, USA \\
	$^{8}$N\'ucleo de Astronom\'ia de la Facultad de Ingenier\'ia, Universidad Diego Portales, Av. Ejercito Libertador 441, Santiago, Chile \\
	$^{9}$Instituto de Astrof\'isica, Facultad de F\'isica, Pontificia Universidad Cat\'olica de Chile, 306, Santiago 22, Chile\\
	$^{10}$Centro de Radioastronom\'ia y Astrof\'isica (CRyA-UNAM), 3-72 (Xangari), 8701, Morelia, Mexico \\
	$^{11}$Departamento de Astrof\'isica, Universidad de La Laguna (ULL), E-38206 La Laguna, Tenerife, Spain \\
	$^{12}$Instituto de Astrof\'isica de Canarias (IAC), E-38205 La Laguna, Tenerife, Spain \\
	$^{13}$Departamento de Astrof\'isica, Facultad de CC. F\'isicas, Universidad Complutense de Madrid, E-28040 Madrid, Spain \\
	$^{14}$Florida Institute of Technology, Melbourne, FL 32901, USA \\
	$^{15}$Department of Astronomy, University of Florida, 211 Bryant Space Science Center, P.O. Box 11205, Gainesville, FL 32611-2055, USA \\
}

\date{Accepted XXX. Received YYY; in original form ZZZ}

\pubyear{2015}

\begin{document}
\label{firstpage}
\pagerange{\pageref{firstpage}--\pageref{lastpage}}
\maketitle

\begin{abstract}
  We present sub-arcsecond 7.5$-$13 \um~imaging- and
  spectro-polarimetric observations of \ngc\ using CanariCam on the
  10.4-m Gran Telescopio CANARIAS. At all wavelengths, we find:
  \emph{(1)} A 90 $\times$ 60 pc extended polarized feature  in the
  northern ionization cone, with a uniform $\sim$44\degr\ polarization
  angle. Its polarization arises from dust and gas emission in the ionization cone, heated by
  the active nucleus and jet, and further extinguished by aligned dust
  grains in the host galaxy. The polarization spectrum of the jet-molecular cloud interaction 
  at $\sim$24 pc from the core is highly polarized, and does not show
  a silicate feature, suggesting that the dust grains are different
  from those in the interstellar medium. 
  \emph{(2)} A southern polarized feature at $\sim$9.6 pc from the core. Its polarization
  arises from a dust emission component extinguished by a large concentration of dust in the galaxy disc. 
  We cannot distinguish between dust emission from magnetically aligned
  dust grains directly heated by the jet close to the core, and
  aligned dust grains in the dusty obscuring material surrounding the
  central engine. Silicate-like grains reproduce the polarized dust
  emission in this feature, suggesting different dust
  compositions in both ionization cones. 
  \emph{(3)} An upper limit of polarization degree of 0.3 per
  cent in the core. Based on our polarization model, the expected
  polarization of the obscuring dusty material is $\lesssim$0.1 per
  cent in the 8$-$13 \um~wavelength range. This low polarization may
  be arising from the passage of radiation through aligned dust grains
  in the shielded edges of the clumps.

\end{abstract}





\section{Introduction}
\label{INTRO}

\ngc\ is the archetypal type 2 active galactic nucleus (AGN), whose
proximity (D = 12.5 Mpc, and 1 arcsec = 60 pc, adopting
H$_{\tiny 0} = 73$ km s$^{-1}$ Mpc$^{-1}$) and high brightness make it
an ideal target for polarimetry. This object is probably the
best-studied AGN regarding polarimetry, and this has allowed us to
obtain a better understanding of the AGN structure. The most important
polarimetric study of \ngc, for the sake of the entire field, was
certainly the detection of polarized broad emission lines in the
optical (0.35$-$0.70 \um) wavelengths by
\citet{Antonucci:1985aa}. This detection is interpreted through
scattering of the radiation from the central engine (black hole and
accretion disc) into our line of sight (LOS) by matter in the
ionization cones. This result allowed us to understand that the
central engine of \ngc\ is obscured by a dusty structure, giving a
major boost to the unified model of AGN. The unified model
\citep[i.e.][]{Lawrence:1991aa,Antonucci:1993aa,Urry:1995aa} of AGN
posits that the observational differences between AGN arise from the
different orientations they present us. In this scheme, the AGN
classification solely depends on the anisotropic obscuration of the
central engine by the optically and geometrically thick distribution
of dust.

Pioneering infrared (IR) polarimetric observations
\citep{Knacke:1974aa,Lebofsky:1978aa} found that the nucleus of \ngc\
was highly obscured and polarized. The observed degree of polarization
rises from 1.6 $\pm$ 0.4 per cent at 1.25 \um~to 2.7 $\pm$ 0.2 per
cent at 3.45 \um, and the position angle (P.A.) of polarization
rotates from 99\degr~$\pm$ 7\degr~at 1.25 \um~to 125\degr~$\pm$
2\degr~at 3.45 \um~in a 6 arcsec~(360 pc) diameter aperture. They
suggested that the IR polarization arises from an obscured power-law
source, which is not directly observable at wavelengths shorter than 1
\um. Using 8$-$13 \um\ observations a the 3.9-m Anglo-Australian
Telescope (AAT), \citet{Aitken:1984aa} found a polarization spectrum
with a uniform degree, 1.39 $\pm$ 0.09 per cent, and a P.A. of
54.8\degr~$\pm$ 1.9\degr\ within apertures of 4.2 arcsec (252 pc) and
5.2 arcsec (312 pc) diameter. Although they were not able to rule out
non-thermal processes, they suggested that dust emission from
non-silicate and featureless grains is responsible for the mid-IR
(MIR) polarization. Using these data and additional 0.36$-$4.8
\um~broad-band imaging polarimetric observations,
\citet{Bailey:1988aa} concluded that the most likely polarization
mechanism at near-IR (NIR) arises from the passage of centrally
produced radiation through aligned dust grains in the obscuring dusty
structure; a mechanism known as dichroic absorption. This result was
further supported by the flip in the P.A. of polarization of
$\sim$70\degr~from NIR to MIR. They interpreted this result as the
change from dichroic absorption to dichroic emission from aligned dust
grains in the dusty structure obscuring the central engine. At shorter
wavelengths, a wavelength-independent intrinsic (after correction for
starlight dilution) polarization of $\sim$16 per cent with a constant
P.A. of $\sim$97\degr~from the ultraviolet
\citep[UV,][]{Antonucci:1994aa} to optical wavelengths
\citep{Antonucci:1985aa} was found. The fact that the P.A. of
polarization at UV and optical is different to that in the NIR,
suggests that different mechanisms of polarization dominate in these
wavelength regimes. Further modeling
\citep{Young:1995aa,Watanabe:2003aa} and observations
\citep{Capetti:1995aa,Packham:1997aa,Lumsden:1999aa,Simpson:2002aa,Lopez-Rodriguez:2015aa}
have shown that electron scattering is the dominant polarization
mechanism from UV to optical, whilst dust distributed within the 2$-$3
arcsec (120$-$180 pc) central region accounts for the polarization
features in the IR \citep{Bailey:1988aa}. Specifically, dichroic
absorption from aligned dust grains within the dusty circum-nuclear
obscurer in \ngc\ is the dominant polarization mechanism in the 1$-$5
\um~wavelength range. Based on these studies, and together with the
lack of evidence of variations in the P.A. of polarization over a
period of months or years, non-thermal polarization mechanisms
(e.g. synchrotron emission from a jet) can be ruled out as a
polarization mechanism in the core of \ngc.
 
At large scales, imaging polarimetric observations
\citep{Scarrott:1991aa} at the V band using the 3.9-m AAT showed that
the optical polarization follows the arm and inter-arm structures of
the galaxy. The degree of polarization increases in the large-scale
inner bar (32 arcsec, 1.92 kpc) at a P.A. of 48\degr, the so-called
NIR bar
\citep{Scoville:1988aa,Schinnerer:2000aa,Emsellem:2006aa}. Further
high-spatial resolution optical (0.5$-$0.6 \um) polarimetric
observations \citep{Capetti:1995aa} using the \textit{Hubble Space
  Telescope} (\textit{HST}) showed a centro-symmetric polarization
pattern along the ionization cones in the $\sim$10 arcsec (600 pc)
central region. This polarization pattern is the signature of a
central point source whose radiation is ultimately polarized through
scattering by dust and/or electrons (and which can be polarized from
multiple scattering into the funnel of the obscuring dusty structure
and/or broad line clouds). At UV and optical wavelengths,
these studies showed that electron scattering is the dominant
polarization mechanism in the ionization cones of \ngc, with the
optical polarization in the off-nuclear regions following the magnetic
field of the NIR bar.

Using high-spatial resolution imaging polarimetric observations at
9.7~\um\ on the 8.1-m Gemini telescope, \citet{Packham:2007aa} found
complex polarization structures within the 2$~\times~$2~arcsec (120 pc
$\times$ 120 pc) central region of \ngc. Specifically, (1)
polarization arising from aligned dust grains in the narrow line
emission regions North of the core, (2) dust being channelled toward
the central engine South, East and West of the core, and (3) a nuclear low, $<$0.6 per cent, polarization interpreted as the polarization arising from the compact ($\le$22 pc)
obscuring dusty structure. Although only one filter was used in this
study, making it difficult to disentangle the several polarization
mechanisms, it showed the potential of MIR polarimetric observations
to study the structure within and around AGN.

In this paper we examine the interaction of the AGN with the
surrounding dusty structures within the central 2 arcsec~(120 pc)
region of \ngc. We also put constraints on the polarization of the
obscuring dusty structure. We performed sub-arcsecond resolution
8$-$13 \um~imaging- and spectro-polarimetric observations using
CanariCam on the 10.4-m Gran Telescopio CANARIAS (GTC) at the Roque de
los Muchachos Observatory. The paper is organized as follows: Section
\ref{OBS_RED} describes the observations and data reduction, Section
\ref{RES} presents our polarimetric results. Polarization models are
developed in Section \ref{PolModel} to explain the several
polarization features, and then analyzed and discussed in Section
\ref{ANA}. In Section \ref{CON} we present the conclusions.


\section{Observations and data reduction}
\label{OBS_RED}


\subsection{Imaging-polarimetry}
\label{OBS_ima}

NGC 1068 was observed on December 29 and 30, 2012, using the imaging
polarimetric mode \citep{Packham:2005aa} of CanariCam
\citep{Telesco:2003aa} on the 10.4-m GTC in Spain. CanariCam uses a
320 $\times$ 240 pixel Si:As Raytheon array, with a pixel scale of
0.0798 arcsec pixel$^{-1}$. The imaging polarimetric mode uses a
half-wave retarder (half-wave plate, HWP), a field mask and a
Wollaston prism. The Wollaston prism and HWP are made with
sulphur-free CdSe. The HWP is chromatic, resulting in a variable
polarimetric efficiency across the 7.5$-$13 \um~wavelength range that
has been well determined \citep{Packham:2008ab}. This mode is usable
across the entire wavelength range. In standard polarimetric
observations, the HWP is set in four P.A. in the
following sequence: 0\degr, 45\degr, 22.5\degr and 67.5\degr. The
field mask consisted of a series of slots of 320 pixels $\times$ 25
pixels each, corresponding to a field of view (FOV) of 25.6 arcsec
$\times$ 2.0 arcsec, where a total of three slots can be used,
providing a non-contiguous total FOV of 25.6 arcsec $\times$ 6.0
arcsec.

The Si2 ($\lambda_{\mbox{\tiny c}}$ = 8.7 \um, $\Delta \lambda =$ 1.1
\um, 50 per cent cut-on/off), Si4 ($\lambda_{\mbox{\tiny c}}$ = 10.3
\um, $\Delta \lambda =$ 0.9 \um, 50 per cent cut-on/off) and Si5
($\lambda_{\mbox{\tiny c}}$ = 11.6 \um, $\Delta \lambda =$ 0.9 \um, 50
per cent cut-on/off) filters provide the largest wavelength coverage
with the best combination of sensitivity and spatial resolution for
the filter set of CanariCam in the 10 \um~atmospheric window. We thus
used this combination of filters. Additionally, to study possible star
formation regions in and around the core of \ngc, we performed
observations using the PAH2 ($\lambda_{\mbox{\tiny c}}$ = 11.3 \um,
$\Delta \lambda =$ 0.6 \um, 50 per cent cut-on/off)
filter. Observations were made using a standard chop-nod technique to
remove time-variable sky background and telescope thermal emission,
and to reduce the effect of 1/{\it f} noise from the
array/electronics. In all observations, the chop-throw was 8 arcsec,
the chop-angle was 0\degr~E of N, and the chop-frequency was 1.93
Hz. The angle of the short axis of the array with respect to North on
the sky (i.e. instrumental position angle, IPA) was 90\degr~E of N,
and the telescope was nodded every 52s along the chopping
direction. Thus, only one slot with a FOV of 25.6 arcsec $\times$ 2.0
arcsec was used in the observations, with the negative images
(produced by the chop-nod technique) within the same slot. For each
filter, two observational sets were observed, except for the PAH2
filter where three observational sets (one observational set the first
night and two observational sets the second night) were observed. A
summary of the observations is shown in Table \ref{table1}. To improve
the signal-to-noise ratio (S/N) of the observations, the negative
images on the array were also used, providing the
total useful on-source time shown in Table \ref{table1}.


\begin{table}
	\caption{Summary of observations.}
	\label{table1}
	\begin{tabular}{ccccc}
		\hline

Date				&	$\lambda$ 	&	On-source time$^{a}$	&	FWHM$_{\mbox{\tiny PSF}}^{b}$	\\
(yyyymmdd)		&	(\um)			&	(s)					&				(arcsec)	\\
					
\hline
	&		&	Imaging-polarimetry \\
\hline	

20121229			&	8.7				&	1456				&	0.38				\\
					&	10.3			&	2040				&	...					\\
		 			&	11.3			&	1020				&	0.39				\\
		 			&	11.6			&	1020				&	...					\\
20121230			&	11.3			&	1894				&	0.39				\\

\hline
	&	&	Spectro-polarimetry \\
\hline
20130722			&	7.5$-$13		&	1894				&	0.35				\\	
\hline
	\end{tabular}\\
	$^{a}$For the imaging-polarimetric observations, the total on-source time was estimated accounting for the positive and negative images, produced by the chop-nod technique, on the array. 
 $^{b}$FWHM of the PSF. 
\end{table}


Data were reduced using custom \textsc{idl} routines. The difference
for each chopped pair was calculated and the nod frames, then
differenced and combined to create a single image per HWP P.A. During
this process, all nods were examined for high or variable background
that could indicate the presence of clouds or variable precipitable
water vapor. Fortunately, no data needed to be removed for these
reasons. As \ngc\ was observed in different sets, each HWP P.A. frame
was registered and shifted to a common position, then images with the
same HWP P.A. were co-added. Next, the ordinary (o-ray) and
extraordinary (e-ray) rays, produced by a Wollaston prism, were
extracted and the Stokes parameters $I$, $Q$ and $U$ were estimated
according to the ratio method \citep[e.g.][]{Tinbergen:2005aa}. Then,
the degree P $= \sqrt{Q^2 + U^2}/I$ and P.A.~$= 0.5\arctan{(U/Q)}$,
of polarization were estimated.

The instrumental polarization was corrected based on data provided by
the GTC website\footnote{Information at
  \url{http://www.gtc.iac.es/instruments/canaricam/canaricam.php\#Instrumental\_Polarization}}. Specifically,
the instrumental polarization is P$_{\mbox{\mbox{\tiny ins}}}$ = 0.6
$\pm$ 0.2 per cent in all filters with a dependence on the P.A. of
polarization given by P.A.$_{\mbox{\tiny ins}}$ = $-(RMA + Elev) -$
29.6\degr, where $RMA$ is the Nasmyth rotator mechanical angle; and
$Elev$ is the telescope elevation. The instrumental polarization was
corrected as follows. The normalized Stokes parameters,
q$_{\mbox{\tiny ins}} = Q_{\mbox{\tiny ins}}/I_{\mbox{\tiny ins}}$ and
u$_{\mbox{\tiny ins}} = U_{\mbox{\tiny ins}}/I_{\mbox{\tiny ins}}$, of
the instrumental polarization were estimated using the degree
P$_{\mbox{\tiny ins}}$ and position angle P.A.$_{\mbox{\tiny ins}}$ of
the instrumental polarization. Then, q$_{\mbox{\tiny ins}}$ and
u$_{\mbox{\tiny ins}}$ were subtracted from the normalized Stokes
parameters of \ngc. Finally, the polarization efficiency was corrected
based on data provided by the GTC website\footnote{Information at: \url{http://www.gtc.iac.es/instruments/canaricam/canaricam.php\#Polarization\_Measurement\_Efficiency}}. Specifically,
the polarization efficiency is 90.3 per cent, 99.5 per cent, 95.9 per
cent, and 97.0 per cent at 8.7, 10.3, 11.6, and 11.3~\um~PAH,
respectively. The measurements of the degree of polarization were
corrected for polarization bias using the approach by
\citet{Wardle:1974aa}.

The polarized young stellar object, AFGL 2403, was observed at 8.7
\um~and 11.3 \um~immediately before \ngc, with an on-source time of
73s in both filters. In total flux, AFGL 2403 was used as point-spread
function (PSF) calibrator.  Table \ref{table1} shows the full width at
half maximum (FWHM) of AFGL 2403, where a Moffat function with two
parameters, FWHM and $\beta$, best described the delivered PSF
\citep[e.g.][]{Radomski:2008aa}. In polarimetry, AFGL 2403 allowed us
to characterize the polarization observations because it is bright,
95.2 Jy, and polarized, 1.5 $\pm$ 0.6 per cent and 1.2 $\pm$ 0.3 per
cent at 8.7 \um~and 11.3 \um, respectively. The degree of polarization
of AFGL 2403 corrected for instrumental polarization and polarization
efficiency at 8.7 \um, was estimated at 1.3 $\pm$ 0.2 per cent. Our
measurement is in excellent agreement with the degree of polarization
of 1.4 $\pm$ 0.8 per cent at 8.7 \um~measured by
\citet[fig. 2]{Smith:2000aa}. The zero-angle of the P.A. of
polarization was calculated as the difference of the P.A. of
polarization from our measurement, $\theta =$ 100\degr $\pm$ 2\degr,
and \citet{Smith:2000aa}, $\theta_{\mbox{\tiny s}} =$ 39\degr $\pm$
4\degr, i.e. $\Delta\theta_{\mbox{\tiny 8.7\um}}$ = $-$61\degr $\pm$
5\degr~at 8.7 \um. At other wavelengths, the zero-angle was estimated
using the wavelength dependence of the P.A. of polarization observed
by \citet[fig. 2]{Smith:2000aa}.

Dedicated flux-standard stars were not observed. Flux calibration was
performed using the N-band spectra observed with Michelle on the 8.1-m
Gemini North \citep{Mason:2006aa}. Specifically, flux calibration was
performed using the spectral points at 8.7, 10.3, 11.3, and 11.6~\um\
of the 0.4 arcsec wide slit oriented 20\degr~E of N and centered at
the peak of \ngc\ \citep[see][fig. 2]{Mason:2006aa}. The extracted
fluxes were 7.5, 8.3, 11.0, and 11.9~Jy, respectively. Then, measured
counts in a 0.4~$\times$~0.4 arcsec simulated slit aperture from our
images at each wavelength were equated to the flux densities from the
MIR spectra by \citet{Mason:2006aa}. Finally, the factor Jy
counts$^{-1}$ was estimated and used in the measurements of the flux
densities in the several apertures shown in Section
$\ref{Res_imaphot}$. These observations can be found at GTC Public
Archive.


\subsection{Spectro-polarimetry}
\label{OBS_specpol}

NGC 1068 was observed on July 22, 2013, during commissioning of the
spectro-polarimetric mode \citep{Packham:2005aa} of CanariCam on the
10.4-m GTC. The spectro-polarimetric mode uses a HWP, a field mask, a
slit and a Wollaston prism. A 0.41 arcsec ($\sim$5 pixels) wide slit
oriented at 0\degr~E of N to cover the North-South polarized features
of NGC 1068 was used. The low resolution N-band
($\lambda_{\mbox{\tiny c}}$ = 10.4 \um, $\Delta \lambda =$ 5.2 \um, 50
per cent cut-on/off) grating was used, resulting in a dispersion of
0.019 \um~pixel$^{-1}$ and R $= \lambda/\Delta\lambda \sim$ 175.

Observations were made using a standard chop-nod technique as
described in Section \ref{OBS_ima} with a chop-throw of 8 arcsec,
chop-angle of 90\degr~E of N and a chop frequency of 1.93 Hz. The IPA
was 0\degr~E of N, and the telescope was nodded every 45.5s along the
chopping direction. Thus, only one slot with a FOV of 25.6 arcsec
$\times$ 2.0 arcsec was used in these observations. We took two
spectro-polarimetric observational sets with an on-source time of 947s
each (Table \ref{table1}). Before these observations, we obtained
acquisition N-band imaging polarimetric observations with an on-source
time of 79s.  The slit was aligned to the peak of the total flux of
NGC 1068 to better than 2 pixels (0.16 arcsec).

Data were reduced as described in Section \ref{OBS_ima}. For each
spectro-polarimetric observational set of \ngc, the o- and e-rays were
extracted in two different locations (Section \ref{RES_specpol}) of
the slit: (1) PSF-extraction of the \emph{Southern feature}, and (2) at 0.35
arcsec (4.4 pixels) North of the peak of the total flux density image
using a fixed 0.4 arcsec (5 pixels) aperture. CanariCam shows a slight
($\sim$1 pixel) curvature across the array in the spectral direction,
which was measured using the observations of the polarized young
stellar object, AFGL 2591. This curvature was taken into account
during the extraction of the spectra. For each observational set,
Stokes parameters, $I$, $Q$ and $U$, were estimated according to the
ratio method, and then co-added to obtain the final spectra. Finally,
the degree and P.A. of polarization were estimated.

The instrumental polarization was corrected as described in Section
\ref{OBS_ima}. The polarization efficiency was corrected using
observations of the unpolarized standard star, HD 186791, through a
wire-grid in the optical path of CanariCam. These observations allow
us to estimate the polarization efficiency spectra of the HWP across
the 7.5$-$13 \um~wavelength range. The polarization efficiency was
corrected by multiplying the spectrum of the degree of polarization of
\ngc\ by the inverse of the polarization efficiency spectrum. Although
the variation of the polarization efficiency was accounted for, a
variable S/N of the Stokes parameters as a function of the wavelength
still remains. We corrected this effect by binning the $Q$ and $U$
spectra across the wavelength axis in bins of 0.285 \um~(15 pixels) to
obtain higher S/N across the spectrum. The measurements of the degree
of polarization were corrected for polarization bias using the
approach by \citet{Wardle:1974aa}.

The polarized young stellar object, AFGL 2591, was observed using the
spectro-polarimetric mode immediately before NGC 1068. In total flux,
AFGL 2591 was used as a PSF calibrator, giving us the FWHM across the
wavelength range for the PSF-extracted spectra. A Moffat function with
two parameters, FWHM and $\beta$, best described the delivered PSF. In
polarimetry, AFGL 2591 allowed us to estimate the zero-angle
calibration of the observations because it is bright, 281.5 Jy, and
highly polarized, 5.6 $\pm$ 1 per cent
\citep[fig. 2]{Smith:2000aa}. The zero-angle calibration,
$\Delta\theta$, was estimated as the difference of the measured
P.A. of polarization of our observations, $\theta =$ 58.5\degr~$\pm$
0.3\degr, and the P.A. of polarization by \citet{Smith:2000aa},
$\theta_{\mbox{\tiny s}} =$ 170\degr~$\pm$ 1\degr. We measured a
constant P.A. of polarization across the 7.5$-$13 \um~wavelength
range, in excellent agreement with \citet{Smith:2000aa}. Thus, the
zero-angle of polarization was estimated to be
$\Delta\theta = \theta_{\mbox{\tiny s}} - \theta =$ 111.5\degr~$\pm$
1.0\degr.

The total flux spectrum of \ngc\ was wavelength-calibrated to an
accuracy of 0.019 \um~pixel$^{-1}$ using 10 \um~window sky lines
present in the spectrum. Then, the total flux spectrum was binned to a
spectral resolution of 0.076 \um~(4 pixels) to improve the S/N. Next,
the spectrum of \ngc\ was divided by the spectrum of the standard
star, HD 186791 (K3III). \citet{Cohen:1995aa} show that late K and M
stars have fundamental vibration-vibration bands of SiO in their
spectra, which significantly depress the spectrum in the 7.5$-$10
\um~wavelength range, and therefore affect the ratio with the science
object. To remove this band and obtain the original spectrum of \ngc,
the total flux spectrum of \ngc\ was multiplied by the template
spectrum of the standard star HD 186791, provided by the Gemini
website\footnote{Template at
  \url{http://www.gemini.edu/sciops/instruments/mid-ir-resources/spectroscopic-calibrations\#waveca}}. Flux
calibration was achieved by using the N-band spectra of \ngc\ (in Jy)
observed with Michelle on the 8.1-m Gemini North \citep{Mason:2006aa}.
The measured nuclear spectrum (in counts) in a 0.4 arcsec aperture
was extracted and equated to the flux density from the MIR
spectrum by \citet{Mason:2006aa}. The factor Jy counts$^{-1}$ was
estimated and used in several extractions of the spectra from our
observations.



\section{Results}
\label{RES}


\subsection{Imaging-polarimetry}
\label{Res_ima}


\subsubsection{Photometry}
\label{Res_imaphot}

\begin{figure*}
	\includegraphics[angle=0,trim=1cm 2cm 0cm 0cm,scale=.16]{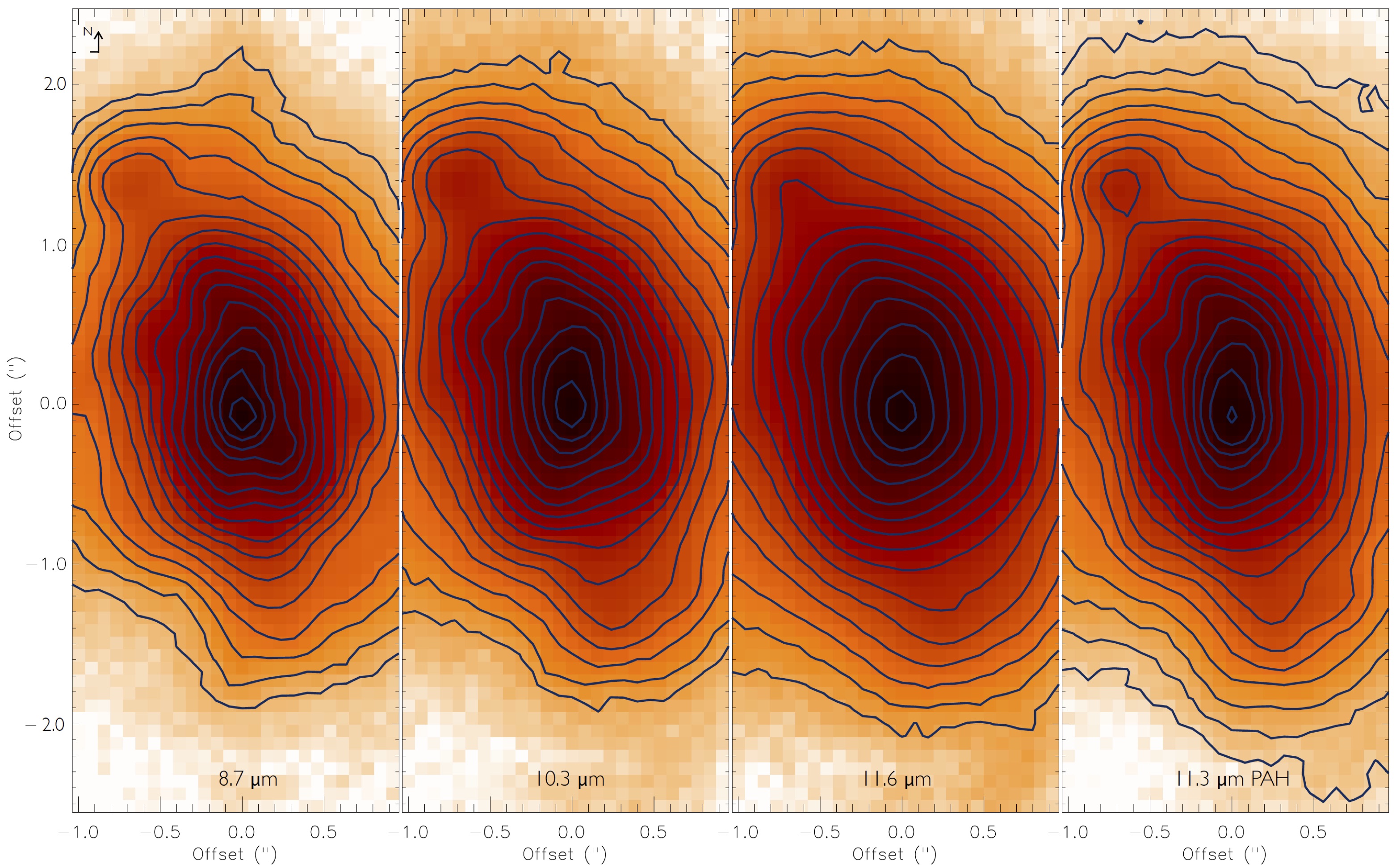}
    \caption{Total flux images of the central 5 arcsec~$\times$~2
      arcsec~(300 pc~$\times$~120 pc) at 8.7 \um, 10.3 \um, 11.6
      \um~and 11.3 \um~PAH. Contours start at 11$\sigma$ and increase
      as 1.5$^{n}$, where $n$ = 7, 8, 9, $\dots$ North is up and East
      is left.}
\label{fig1}
\end{figure*}

Fig. \ref{fig1} shows the total flux images at 8.7, 10.3, 11.6, and
11.3~\um\ (PAH). At all wavelengths \ngc\ appears as an unresolved
core in the E--W direction, with extended emission along N--S. The
N--S extended emission slightly bends to the N--W direction up to 0.5
arcsec (30 pc), and then to the N--E direction up to 1.4 arcsec (84
pc); this feature was termed as \emph{`tongue'}
\citep[see][fig. 2]{Bock:2000aa}. At 1.4 arcsec N--E from the core,
the \emph{NE knot} is located. These structures are in excellent
agreement with previously published high-spatial resolution MIR
observations
\citep[e.g.][]{Bock:2000aa,Tomono:2001aa,Galliano:2005aa,Poncelet:2007aa}.


We made measurements of the nuclear total flux density images centred
at the peak of the total flux density at several diameter apertures
(hereafter, aperture refers to diameter) to compare with previously
published values (Table \ref{table2}). In all cases, photometric
errors were estimated by the variation of the counts in subsets of the
data. Our photometric measurements are consistent with published flux
density measurements. We also made photometric measurements of the
\emph{Southern feature} and \emph{North knot} (Section
\ref{Res_imapol}, Fig. \ref{fig2} and \ref{fig3}) in a 0.4 arcsec (24
pc) aperture centered on the peaks of polarized flux, for all
features, and at each wavelength (Fig. \ref{fig2} and \ref{fig3}).


\begin{table}
\caption{Comparison of the nuclear total flux density of NGC 1068 with literature.}
\label{table2}

\begin{tabular}{cccc}

\hline
$\lambda$			&	Aperture$^{\mbox{\tiny 1}}$			&	Flux density			&	Ref(s).		\\
(\um)				&	(arcsec)				&	(Jy)					&					\\
		
\hline

8.7					&	2					&	$15 \pm 2$ 				&	a					\\
					&	3					&	$14.8 \pm 1.5$			&	b 	\\
					&	3.8				&	$15 \pm 1$				&	b	     	\\	
10.3				&	2					&	$19 \pm 3$				& 	a	     				\\
					&	3.8				&	$20 \pm 2$				& 	b	     	\\
11.3				&	0.4				&	$16 \pm 2$				&	a		     			\\
					&	0.4				&	$11.5 \pm 1.5$			&	c		     	\\
11.6				&	2.0				&	$33 \pm 	5$				&   a		     			\\
					&	1.2				&	$30 \pm 5$				&	d	     	\\	
					&	1.8				&	$39 \pm 	6$				&	d	     	\\	
					&	2.0				&	$40 \pm 	6$				&	d	     	\\			
\hline
\end{tabular}
\\
Notes: $^{\mbox{\tiny 1}}$Aperture refers to diameter.
References: (a) This work; (b) \citet{Tomono:2006aa}; (c) \citet{Jaffe:2004aa}; d) 11.6 \um~imaging observations by \citet{Mason:2006aa}\\
\end{table}



\subsubsection{Polarimetry}
\label{Res_imapol}

\begin{figure*}
\includegraphics[angle=0,trim=0.5cm 1.5cm 0cm 0cm,scale=.15]{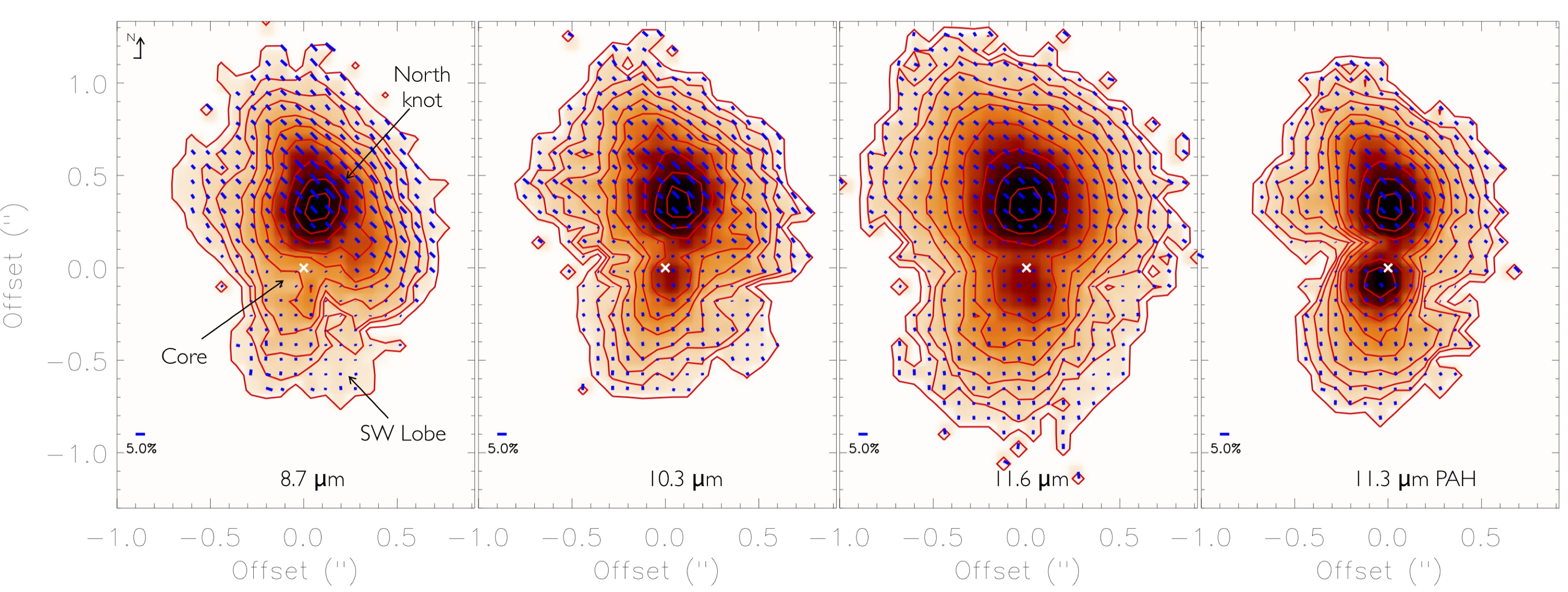}
\caption{Polarized flux images of the central 3~$\times$~2~arcsec
  (180~$\times$~120~pc) at 8.7, 10.3, 11.6, and 11.3~\um\ (PAH) with
  the overlaid polarization vectors. Contours are plotted in steps of
  10 per cent from the peak pixel of the \emph{North knot} where only those
  polarization vectors with $P/\sigma{\mbox{\tiny p}} \ge 3 $ are
  shown. A vector of 5 per cent of polarization is shown for
  scale. The white crosses show the location of the peak of the total
  flux images (Fig. \ref{fig1}). North is up and East is left.}
\label{fig2}
\end{figure*}

\begin{figure*}
\includegraphics[angle=90,trim=0cm 1cm 0cm 0cm,scale=.25]{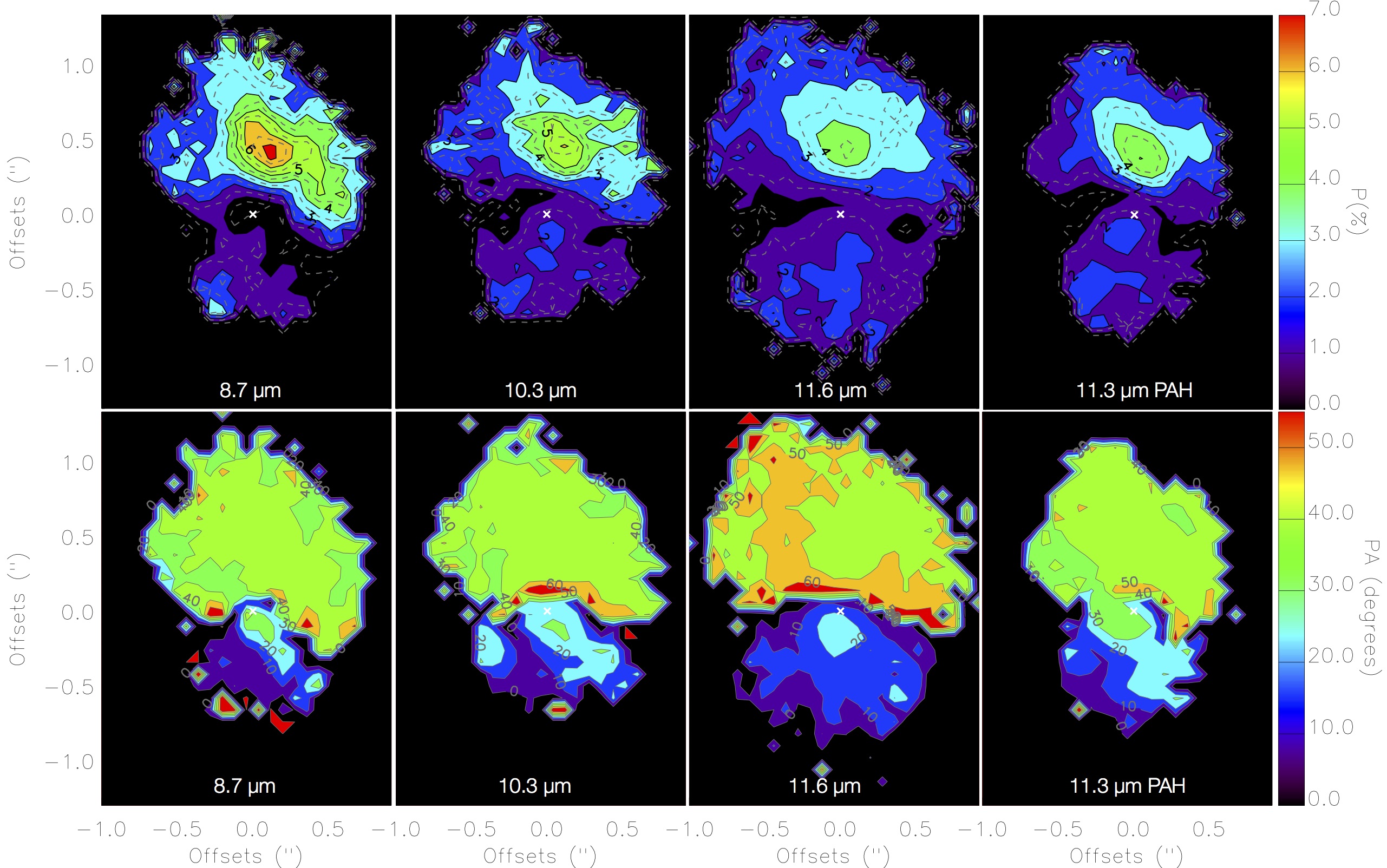}
\caption{Contours of the degree (first row) and P.A. (second row) of
  polarization of the central 3~$\times$~2~arcsec
  (180~$\times$~120~pc) region at 8.7, 10.3, 11.6, and 11.3~\um\
  (PAH). For the degree of polarization, contours start at 0 per cent
  and increase in steps of 0.5 per cent. For the P.A. of polarization,
  contours are plotted in steps of 10\degr. The white crosses show the
  location of the peak of the total flux images
  (Fig. \ref{fig1}). North is up and East is left.}
\label{fig3}
\end{figure*}

Fig. \ref{fig2} shows the polarized flux images with the overlaid
polarization vectors at 8.7, 10.3, 11.6 and 11.3~\um\ (PAH). In these
figures, the overlaid polarization vectors are proportional in length
to the degree of polarization, and their orientation shows the P.A. of
polarization. Only polarization vectors with
$P/\sigma{\mbox{\tiny p}} \ge 3 $ are shown. $\sigma{\mbox{\tiny p}}$
is the uncertainty of the degree of polarization per pixel, estimated
as
$\sigma{\mbox{\tiny p}} = \sqrt{\sigma{\mbox{\tiny q}}^2 +
  \sigma{\mbox{\tiny u}}^2}$,
where $\sigma{\mbox{\tiny q}}$ and $\sigma{\mbox{\tiny u}}$ are the
uncertainties of the normalized Stokes parameters
\citep{Wardle:1974aa}. Fig. \ref{fig3} shows the contours of the
degree and P.A. of polarization for those polarization vectors with
$P/\sigma{\mbox{\tiny p}} \ge 3 $. At all wavelengths, several
polarized features are found:

\begin{enumerate}
\item The most prominent feature is the uniform P.A. of polarization
  with an extension of $\sim$1.5 arcsec $\times$ 1 arcsec (90 pc
  $\times$ 60 pc) North of the peak of the total flux density (white
  crosses in Fig. \ref{fig2}). The P.A. of polarization is roughly
  constant, $\sim$44\degr, in an area of 1.4 arcsec~$\times$~0.8
  arcsec (84 pc~$\times$~48 pc), with variations of $\sim$10\degr\
  (Fig. \ref{fig3}, bottom). In this region, we found a resolved
  polarized knot at $\sim$0.4 arcsec (24 pc) North from the peak of
  the total flux images, termed \emph{`North knot'}. For the \emph{North
  knot}, the degree of polarization shows a steep increase from South
  to North, with a peak in the degree of polarization, which then
  decreases smoothly to the NW and SW directions
  (Fig. \ref{fig3}-up). The degree of polarization decreases at longer
  wavelengths (Table \ref{table4}, Fig. \ref{fig4}) with a constant
  P.A. of polarization across the wavelength range.

\item The polarized flux images (Fig. \ref{fig2}) show a peak slightly
  shifted, $\sim$0.16 arcsec (9.6 pc), SE from the peak of the total
  flux images. The degree of polarization increases at longer
  wavelengths with a P.A. of polarization increasing from
  12\degr~$\pm$~10\degr~to 37\degr~$\pm$~5\degr~at 8.7 \um~and 11.6
  \um, respectively.

\item At $\sim$0.5 arcsec (30 pc) South from the peak of the total
  flux images, an extended polarized feature is found. This structure
  is spatially coincident with the \emph{`SW Lobe'} observed in
  polarized flux at 2.0 \um~using {\it HST}/NICMOS by
  \citet{Simpson:2002aa} and in the K$'$ band using MMT/MMT-Pol by
  \citet{Lopez-Rodriguez:2015aa}.

\end{enumerate}



We made measurements of the nuclear polarization to compare with the
literature (Table \ref{table3}). In all cases, polarimetric errors
were estimated by the variation of the counts in subsets of the
data. Our measured nuclear polarization in a 2 arcsec~(120 pc)
aperture at all wavelengths is higher than the degree of polarization
of 1.30~$\pm$~0.05 per cent in a 2 arcsec~aperture in the N-band using
the 3.9-m AAT \citep{Lumsden:1999aa}. These authors measured a P.A. of
polarization of $49\degr~\pm~3\degr$, which is in marginal agreement
with our measurements within the uncertainties. We interpret these
results as (1) an increase in the measured degree of polarization by
an improvement in the spatial resolution, (2) the mix of different
polarization structures within the aperture, and (3) the use of a wide
bandwidth within the 10 \um~atmospheric window, which implies that
their measurements were more strongly affected by the wavelength
dependence of the different mechanisms of polarization. We note that
the P.A. of polarization by \citet{Lumsden:1999aa} is in better
agreement with our measured P.A. of polarization in the Northern
ionization cones (Table \ref{table4}). This result indicates that
their nuclear polarization measurement was dominated by the
polarization arising from the Northern ionization cone (Section
\ref{ANA_North}), as a larger
amount of polarized flux is coming from the North regions within their
2 arcsec~(120 pc) aperture. Within the nuclear 1.7~$\times$~1.2 arcsec
(102~$\times$~72 pc) elliptical aperture, our measured degree of
polarization of 2.2~$\pm$~0.7 per cent at 8.7 \um~is in excellent
agreement with the measured degree of polarization of 2.48~$\pm$~0.57
per cent at 9.7 \um~using the 8.1-m Gemini \citep{Packham:2007aa}. We
note a difference in the measured P.A. of polarization of
26.7\degr~$\pm$~15.3\degr~at 9.7 \um~by \citet{Packham:2007aa}, and
our measured 46\degr~$\pm$~9\degr~and 53\degr~$\pm$~8\degr~at 8.7
\um~and 10.3 \um,~respectively. Both measurements are in marginal
agreement within the uncertainties, however we note that the P.A. of
polarization by \citet{Packham:2007aa} is in better agreement with our
measurements of the southern polarized feature (Table
\ref{table4}). This difference can be understood as the polarimetric
observations of \citet{Packham:2007aa} were more sensitive to the
central $\sim$1 arcsec due to their low S/N. Our degree and P.A. of
polarization are calibrated to standard stars for polarization
measurements, but we note that the total and polarized flux
morphologies of the Gemini/MICHELLE polarimetric data by
\citet{Packham:2007aa} seem to be rotated by 180\degr~from our
dataset.  We confirmed the CanariCam orientation through comparing our
morphology with other total flux measurements, including de-convolved
images, and find a high degree of spatial coincidence.  We speculate
that perhaps the world coordinate system (WCS) used to indicate North
in the Gemini/MICHELLE polarimetric data by \citet{Packham:2007aa} was
incorrect, perhaps due to an additional reflection or translation in
the polarimetry mode compared to direct imaging.

Based on the observed polarized features and their S/N, we made
polarimetric measurements of (1) the \emph{North knot} in a 0.4 arcsec
aperture, and (2) the \emph{Southern feature} at 0.16 arcsec (9.6 pc) to the
peak of the total flux density in a 0.4 arcsec (24 pc) aperture. The
total flux density, degree of polarization, P.A. of polarization and
polarized flux density for each feature and filter are shown in Table
\ref{table4}. Figure \ref{fig4} shows the measurements (blue squares)
for each feature, as well as the used apertures (blue circles in the
central panel). We found an aperture dependence on the polarization in
the Northern ionization cone. Specifically, a constant degree, $\sim$5
per cent, and P.A., $\sim44\degr$, of polarization is measured using a
1.4~$\times$~0.8 arcsec (84~$\times$~48 pc) elliptical aperture with
P.A. $=45\degr$.
 

\begin{table}
  \caption{Comparison of the nuclear polarization of \ngc\ with literature.}
  \label{table3}

\begin{tabular}{ccccc}

\hline
Aperture			&	$\lambda$		&	P					&	P.A.					&	Ref(s).		\\
(arcsec)				&	(\um)				&	(per cent)					&	(\degr)				\\
		
\hline

$1.7 \times 1.2$	 	& 8.7					&	$2.2 \pm 0.7$			&	$46 \pm 9$	  		& 	a \\
							& 9.7					&	$2.48 \pm 0.57	$     	&	$26.7 \pm 15.3$	&	b	\\
							& 10.3		 		&	$1.8 \pm 0.3$			&	$53 \pm 8$	  		&	a\\	
							& 11.3				&	$1.6 \pm 0.3$			&	$52 \pm 5$	  		&	a	\\
							& 11.6				&	$1.7 \pm 0.3$			&	$59 \pm 5$	  		&	a\\	
$2$		 				& 10					&	$1.30 \pm 0.05$		&	$49 \pm 3$			&	c	\\
\hline
\end{tabular}
\\
References: (a) This work; (b) \citet{Packham:2007aa}; (c) \citet{Lumsden:1999aa} \\
\end{table}



\begin{table}
\caption{The measured flux density, degree and P.A. of polarization and polarized flux for the \emph{Southern feature} and \emph{North knot}.}
\label{table4}

\begin{tabular}{ccccc}

\hline
	$\lambda$		&	F$^{(a)}$		&	P					&	P.A.				&	F $\times$ P			\\
		(\um)				&	(Jy)					&	(per cent)					&	(\degr)		&	(mJy)		\\
		
\hline
& & \emph{Southern feature}	\\
\hline
8.7	&	$7.0	\pm	0.8$		&	$0.9	\pm	0.3$		&	$12	\pm	10$	&	$63	\pm	32$	\\
10.3	&	$7.2	\pm	0.8$		&	$2.1	\pm	0.3$		&	$33	\pm	7$		&	$152	\pm	41$	\\
11.3	&	$10.4	\pm	1.2$		&	$2.0	\pm	0.2$		&	$35	\pm	3$		&	$207	\pm	47$	\\
11.6	&	$11.2	\pm	1.1$		&	$1.9	\pm	0.2$		&	$37	\pm	5$		&	$213 \pm	46$	\\
\hline
& & \emph{North knot}	 \\
\hline
8.7	&	$4.5	\pm	0.4$		&	$7.0	\pm	0.6$		&	$40	\pm	1$	&	$317	\pm	48$	\\
10.3	&	$6.1	\pm	0.6$		&	$6.0	\pm	0.4$		&	$45	\pm	1$	&	$366	\pm	57$ \\
11.3	&	$7.2	\pm	0.7$		&	$5.0	\pm	0.4$		&	$50	\pm	1$	&	$362	\pm	67$	\\
11.6	&	$8.4	\pm	0.7$		&	$4.5	\pm	0.3$		&	$45	\pm	1$	&	$378	\pm	59$	\\
\hline	
\end{tabular}
\\
$^{(a)}$Flux densities estimated in a 0.4 arcsec (24 pc) aperture centred in the \emph{Southern feature} and \emph{North knot}, respectively.
\end{table}



\subsection{Spectro-polarimetry}
\label{RES_specpol}


\begin{figure*}
\centering
\includegraphics[angle=0,trim=0cm 0.5cm 0cm 0.5cm,scale=.30]{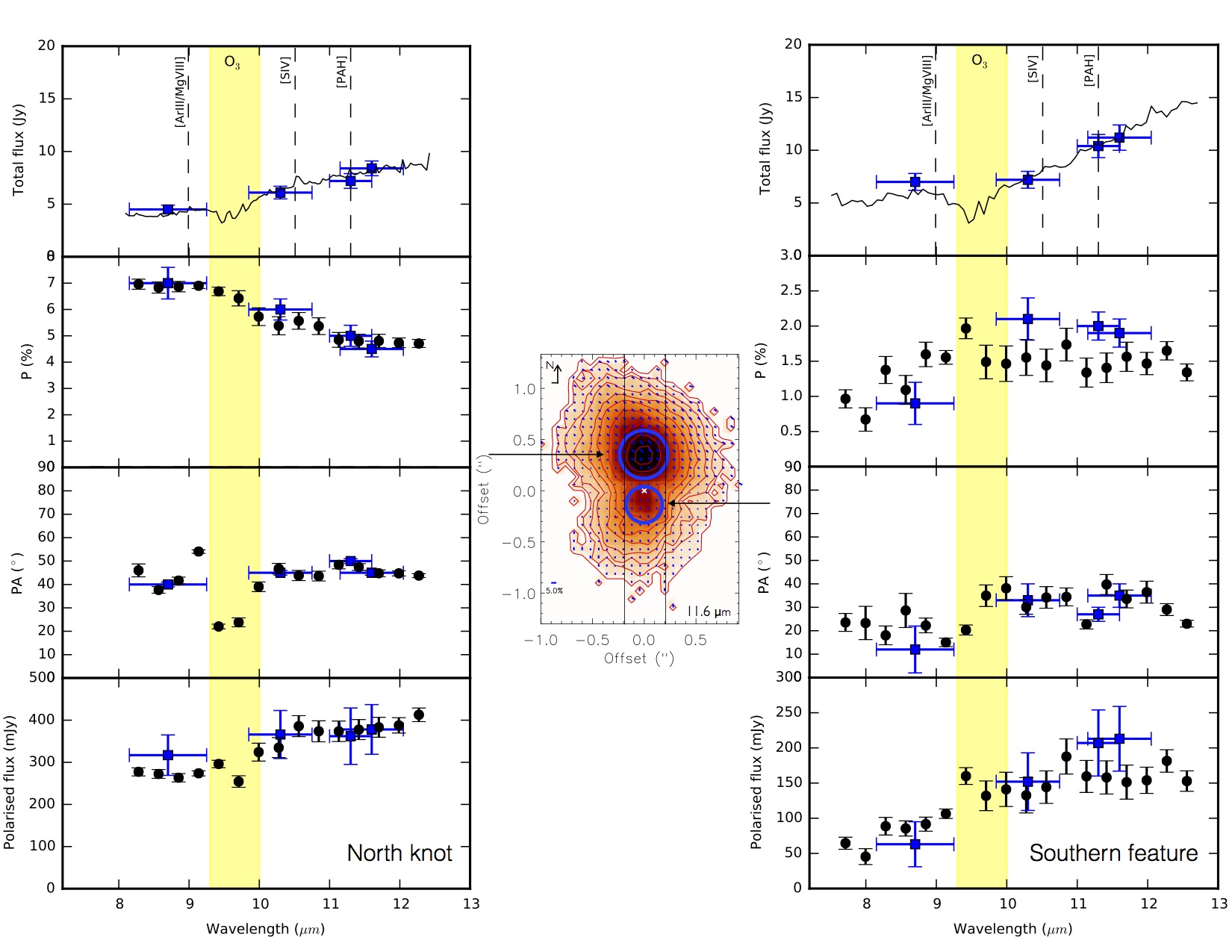}
\caption{The measured total flux density and
  polarization of the \emph{North knot} (left) and \emph{Southern feature}
  (right). The polarized flux image (center) with overlaid
  polarization vectors at 11.6 \um~as in Fig. \ref{fig2} is shown. The
  blue circles in the central panel represent the aperture used for
  the imaging-polarimetric measurements (Section \ref{Res_imapol}),
  and the two vertical black solid lines demarcate the position and
  width of the 0.41 arcsec ($\sim$5 pixels) slit. The total flux
  density (first row) spectra (black solid line) and imaging (blue
  squares) measurements are shown with a spectral bin of 0.076 \um~(4
  pixels). The dashed lines indicate the positions of the
  [\ion{Ar}{iii}/\ion{Mg}{vii}] blend at 8.99 \um, [\ion{S}{iv}]
  fine-structure at 10.51 \um, and 11.3 \um~PAH emission lines, whilst
  the yellow region shows the approximate extent of the telluric
  O$_{3}$ band. The degree (second row), P.A. (third row) of
  polarization and polarized flux density (fourth row) of the imaging-
  (blue squares) and spectro-polarimetric (black circles) observations
  are shown. The spectro-polarimetric observations have a spectral bin
  of 0.285 \um~(15 pixels).}
\label{fig4}
\end{figure*}


Fig. \ref{fig4} shows the total flux density (black solid line), the
degree and P.A. of polarization and the polarized flux density (black
filled-dots) for the PSF-extracted spectra of the 0.4 arcsec (24 pc)
fixed aperture spectra of the \emph{North knot} (left), and the
\emph{Southern feature} (right). The 0.41 arcsec ($\sim$5 pixels) slit
(black solid vertical lines) is shown. The total flux spectra were
binned to a 0.076 \um~(4 pixels) resolution, whilst a 0.285 \um~(15
pixels) bin was used for the degree, P.A. of polarization and
polarized flux. The uncertainties were estimated as a variation of the
counts within the binned data.

For the \emph{Southern feature}, the total flux spectrum shows strong dust
continuum emission with a broad silicate absorption feature. This
spectrum is similar and consistent with the 0.4~$\times$~0.4~arcsec
(24~$\times$~24~pc) slit aperture of the 0.4 arcsec SSW spectrum using
the 8.1-m Gemini by \citet{Mason:2006aa}. The degree of polarization
slightly increases with increasing wavelength. The P.A. of
polarization slightly increases from 12\degr~$\pm$ 10\degr~to
37\degr~$\pm$ 5\degr~across the 8$-$12 \um~wavelength range. The
polarized flux increases with wavelength.

For the \emph{North knot}, the total flux spectrum is similar to and
consistent with the 0.4 arcsec NNE spectrum of \citet{Mason:2006aa},
also showing strong dust continuum emission with a broad silicate
absorption feature and narrow fine-structure lines. In this spectrum,
a [\ion{Ar}{iii}/\ion{Mg}{vii}] blend at 8.99 \um, and a [\ion{S}{iv}]
fine-structure at 10.51 \um~are detected. The degree of polarization
decreases with increasing wavelength and the P.A. of polarization is
roughly uniform through the spectrum. The polarized flux increases
with increasing wavelength.

\citet{Aitken:1984aa} measured a uniform degree, 1.29~$\pm$~0.09 per
cent, and P.A., 54.8\degr~$\pm$~1.9\degr, of polarization across the
8$-13$ \um~wavelength range using a 4.2 arcsec (252 pc) beam. We note
that their measured P.A. of polarization is consistent with the
P.A. of polarization of the northern ionization cone measured in our
observations. We thus gather that the measurements of
\citet{Aitken:1984aa} were dominated by the northern ionization cone
within their 4.2 arcsec (252 pc) beam, as a larger amount of polarized
flux is coming from the North regions within their beam. This result
makes it difficult to obtain information about the core of NGC 1068,
and shows the potential of subarcsecond resolution polarimetric MIR
observations to obtain more sensitive observations with the aim to
study the cores of AGNs.


\section{Polarization model}
\label{PolModel}

We aim to reproduce the observed polarization in the 7.5-13 \um~wavelength
range of the several observed features discussed in Section
\ref{RES}. Specifically, we wish to reproduce the observed degree and
position angle of polarization for 1) the northern and southern ionization cones, and 2) the obscuring dusty structure. 

\subsection{Components of the polarization model}
\label{PolModel_def}

MIR spectro-polarimetric observations of 55 objects (e.g. young
stellar objects, star formation regions, and active galactic nuclei)
using 4-m class telescopes found that 90 per cent of the objects can
be explained by dichroism \citep{Smith:2000aa}. Dichroic extinction
and emission can compete at a given wavelength. Both mechanisms need
to be disentangled through a multi-wavelength study. Fortunately, some
dust grain features, such as silicates, are present in the MIR
wavelength range, and can be used to distinguish between the two
mechanisms. In the case of silicates, the absorptive polarization
shows a peak at $\sim$10 \um, while the emissive polarization
is less structured. In general, if a rotation of the polarization
angle with wavelength is observed, then more than one mechanism of
polarization may be present. Unless the several polarization
mechanisms have intrinsically the same polarization angle, the net
polarization angle will be a function of wavelength. Both the degree
and angle of polarization profiles are crucial to investigate the
emissive and absorptive polarization components at MIR wavelengths.

The dichroic components can be disentangled following the procedure put forward by
\cite{Aitken:2004aa}. These authors presented a procedure to separate
and identify the absorptive, p$_{\mbox{\tiny a}}$, and emissive,
p$_{\mbox{\tiny e}}$, polarization components within the 7.5$-$13
\um~wavelength range. The emissive polarization component is given by
p$_{\mbox{\tiny e}} =$
$|$p$_{\mbox{\tiny a}} / \tau_{\tiny \lambda}|$, where
$\tau_{\tiny \lambda}$ is the wavelength-dependent extinction
curve. This condition can be used as long as the difference in
orthogonal optical depths of the dust grains is less than unity for
p$_{\mbox{\tiny a}}$, which makes p$_{\mbox{\tiny a}}$ independent of
the optical depth. This situation will hold if
$\tau_{\mbox{\tiny MIR}}<$~a few tens. For the extinction curve, the
standard silicate-graphite interstellar dust grain of
R$_{\mbox{\tiny v}} = 5.5$ is assumed\footnote{Note that recent
  studies \citep*{Gao:2013aa} suggested that a trimodal grain size
  distribution with a combination of R$_{\mbox{\tiny v}} = 2.1$, $3.1$
  and $5.5$ is required to achieve a good fit to the extinction curve
  of the Galactic center from the UV to the MIR.}, which appears to be
in agreement with the MIR extinction at different sightlines through
the Galactic center \citep{Weingartner:2001aa}. Variations in our
model of less than 2\% are found when extinction curves of
R$_{\mbox{\tiny v}} = 2.1$, and $3.1$ were used. For the absorptive
component, p$_{\mbox{\tiny a}}$, we took the Becklin-Neugebauer (BN)
object in Orion \citep{Aitken:1989aa} because it is the best-defined
absorptive component with the highest S/N, and commonly used in these studies \citep{Smith:2000aa}.

The polarization model to separate the absorptive and emissive
polarization components is constructed as follows. It assumes an
emission source that can be either polarized or unpolarized, and which
is viewed through a cold dichroic sheet of dust grains. The Stokes
parameters from both components simply add up, and the observed Stokes
parameters, $q_{\mbox{\tiny obs}}$ and $u_{\mbox{\tiny obs}}$, can be
fitted as a linear combination of both components:
\begin{equation}
q_{\mbox{\tiny obs}} = q_{\mbox{\tiny a}} + q_{\mbox{\tiny e}} = A\mbox{p}_{\mbox{\tiny a}} + B\mbox{p}_{\mbox{\tiny e}} \\
\\
u_{\mbox{\tiny obs}} = u_{\mbox{\tiny a}} + u_{\mbox{\tiny e}} = C\mbox{p}_{\mbox{\tiny a}} + D\mbox{p}_{\mbox{\tiny e}} 
\end{equation}
\noindent
where $A$, $B$, $C$ and $D$ are the constants used for the fitting
procedure. We followed the fitting procedure described by
\cite{Aitken:2004aa}. Then, the final degree of polarization,
P$_{\mbox{\tiny obs}} = \sqrt{q_{\mbox{\tiny obs}}^2 + u_{\mbox{\tiny
      obs}}^2}$,
and the position angle,
P.A$._{\mbox{\tiny obs}} = 0.5\arctan{(u_{\mbox{\tiny
      obs}}/q_{\mbox{\tiny obs}})}$, are estimated.

The total flux spectrum is defined as an extinguished blackbody
component with a characteristic temperature T$_{\mbox{\tiny BB}}$. We
have assumed an extinction curve with R$_{\mbox{\tiny v}} = 5.5$, and
the characteristic temperature was sampled in steps of 10 K. The polarized flux was estimated
as the modeled total flux, times the total degree of polarization,
P$_{\mbox{\tiny obs}}$.

\subsection{Fitting of the ionization cones}
\label{PolModel_Fit1}

In the case of the northern ionization cone, a constant P.A. of
polarization (Table \ref{table4}, Fig. \ref{fig4}-left) suggests a
unique polarization mechanism. In objects in our Galaxy where MIR
polarization arises through dichroic absorption by aligned dust
grains, a prominent silicate peak in the degree of polarization is
present just longwards of 10
\um~\citep[i.e.][]{Aitken:1986aa,Smith:2000aa}. No such feature is
present in the polarization spectrum of the \emph{North knot} in \ngc, which
instead shows a slow decrease in the degree of polarization with
increasing wavelength. Thus, non-silicate features, i.e. non 9.7 \um~feature, are present in the MIR
polarization. We give a further interpretation of these features in
Section \ref{ANA_North}.


\begin{figure}
\centering
\includegraphics[angle=0,trim=1cm 0.5cm 0cm 0cm,scale=0.27]{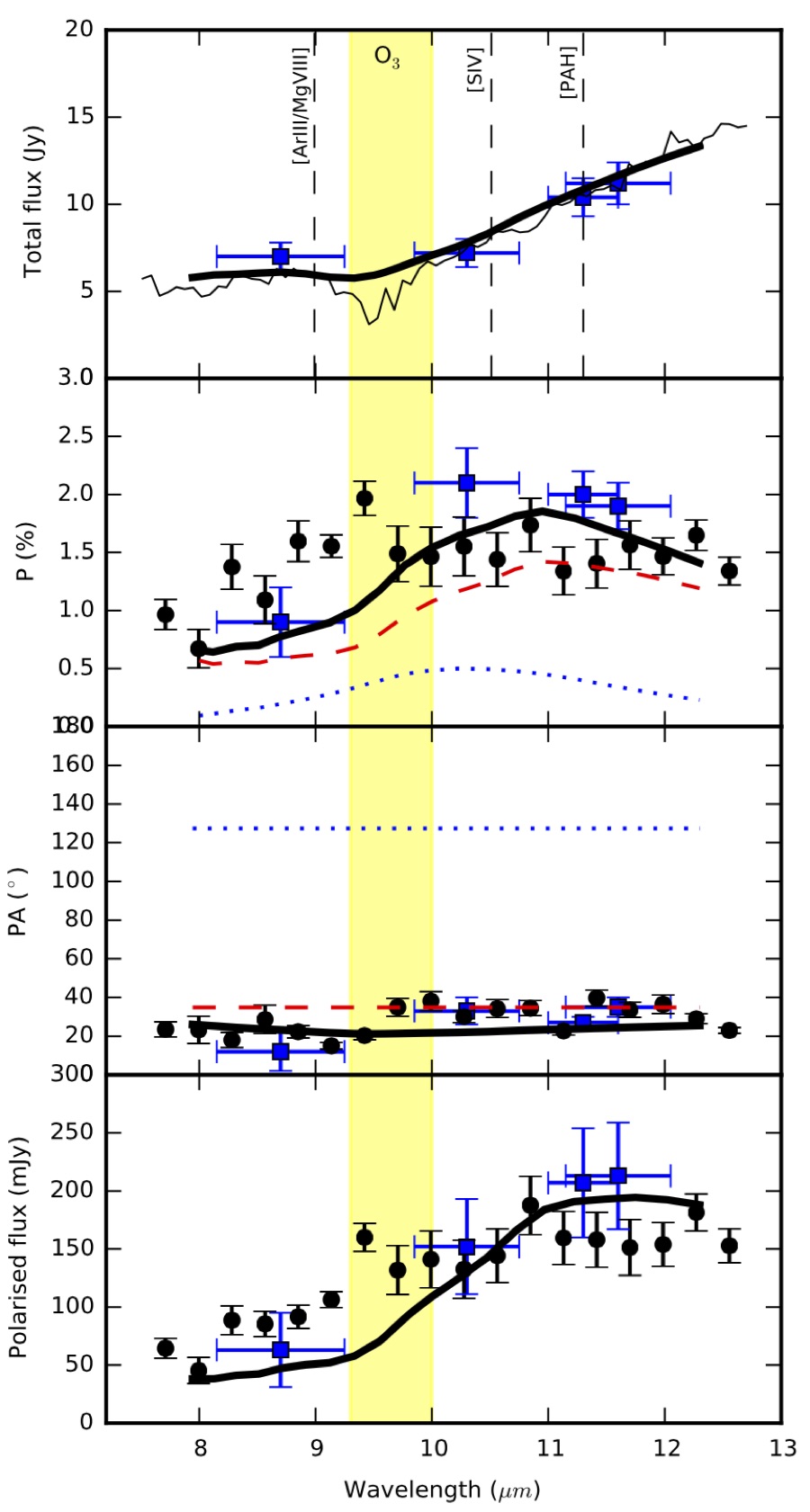}
\caption{\emph{Southern feature} measurements as in
  Fig. \ref{fig4}, with the total model (black solid line), and
  the absorptive (blue dotted line) and emissive (red dashed line)
  polarization components from our polarization model presented in
  Section \ref{ANA_South}.}
\label{fig5}
\end{figure}


In the southern ionization cone, the rotation of the P.A. of
polarization (Table \ref{table4}, Fig. \ref{fig4}-right) suggests that
more than one mechanism contributes to the polarization. Following the procedure described in Section \ref{PolModel_def}, we simultaneously fit the measured total flux, degree and P.A. of
polarization of the \emph{Southern feature} (Figure \ref{fig5}). The
polarization model has four free model parameters: 1) the visual
extinction A$_{\mbox{\tiny v}}$ towards the core, 2) the blackbody
characteristic temperature T$_{\mbox{\tiny BB}}$, 3) the degree of
polarization, and 4) the P.A. of polarization. The fit was considered
acceptable when the reduced $\chi^{2}$ was minimized. We obtain
a reduced $\chi^{2}$ = 0.21, 2.22, 15.8 and 14.2 for the flux density,
degree and P.A. of polarization, and polarized flux density,
respectively. Although we are aware of the limitations of the
$\chi^{2}$ fitting procedure, we are more interested on explaining the
overall behaviour of the polarization spectrum rather than optimize
the goodness of the fitting, as we are limited by the low S/N of the
observations. Further development of a Bayesian fitting procedure is
out of the scope of this paper, but we think it will benefit the
statistical analysis presented here as shown by
\citet{Lopez-Rodriguez:2016aa}. For the best fit, we found a blackbody
component with a characteristic temperature of
T$_{\mbox{\tiny BB}} = 210 \pm 10$~K, extinguished with
A$_{\mbox{\tiny v}} =$ 9 $\pm$ 1 mag. The emissive polarization
component is dominant and describes the observed wavelength dependence
of the degree of polarization with a P.A. = 35\degr~$\pm$ 1\degr. The
absorptive component contributes $<$ 20 per cent of the polarization
with a P.A. = 127\degr~$\pm$ 1\degr. We give a further interpretation
of this feature in Section \ref{ANA_South}.

\subsection{Fitting of the dusty central structure}
\label{PolModel_torus}

We now investigate the expected MIR polarization of the obscuring
dusty structure in terms of dichroism. Dichroic absorption is the
dominant mechanism of polarization at NIR wavelengths in the core of
\ngc. This allows us to constrain the absorptive and emissive
polarization components in the 2$-$13 \um~wavelength range. For the
absorptive component, we assume a Serkowski \citep{Serkowski:1975aa}
curve up to 8 \um, followed by the absorptive profile of the BN object
in Orion (Section \ref{ANA_South_feature}) in the 8-13 \um~wavelength
range. This composition takes into account the silicate feature at MIR
wavelengths, that, if not used, would lead to the expected MIR
polarization being underestimated by a factor $>$100. Using NIR
polarimetric adaptive optics observations,
\citet{Lopez-Rodriguez:2015aa} estimated an intrinsic
(starlight-corrected) polarization at K$'$ of 7.0 $\pm$ 2.2 per cent
in a 0.5 arcsec (30 pc) aperture. Using this value to constrain the
absorptive polarization allows us to estimate the intrinsic
polarization of the obscuring dusty structure in \ngc. To constrain
the extinction curve (with R$_{\mbox{\tiny v}} =$ 5.5 as in
Sec.~\ref{ANA_South_feature}), lower and upper limits of the visual
extinction towards the core of \ngc\ must be assumed. For the upper
limit we used A$_{\mbox{v}} =$ 39 mag
\citep{Packham:1997aa,Watanabe:2003aa,Lopez-Rodriguez:2015aa},
interpreted as the extinction to the NIR emitting regions. For the
lower limit, A$_{\mbox{v}}$ is estimated using the 9.7~\um\ silicate
feature, $\tau_{\mbox{\tiny 9.7\um}} \sim$ 0.41, of the nuclear
spectrum in a 0.4~$\times$~0.4 arcsec (24$\times$~24 pc) slit aperture
by \citet{Mason:2006aa}. This result was then converted\footnote{The
  conversion of the 9.7 \um~silicate feature to visual extinction is
  A$_{\mbox{\tiny v}} = 18.5\,\tau_{\mbox{\tiny 9.7\um}}$
  \citep{Roche:1984aa}} to A$_{\mbox{v}} \sim$ 8 mag. The difference
of both visual extinctions can be interpreted as a temperature
gradient and inhomogeneities (clumpiness) in the obscuring dusty
material \citep[i.e.][]{Pier:1992aa,Imanishi:2000aa,Levenson:2007aa}.

\begin{figure}
\centering
\includegraphics[angle=0,trim=0cm 1cm 0cm 0cm,scale=0.35]{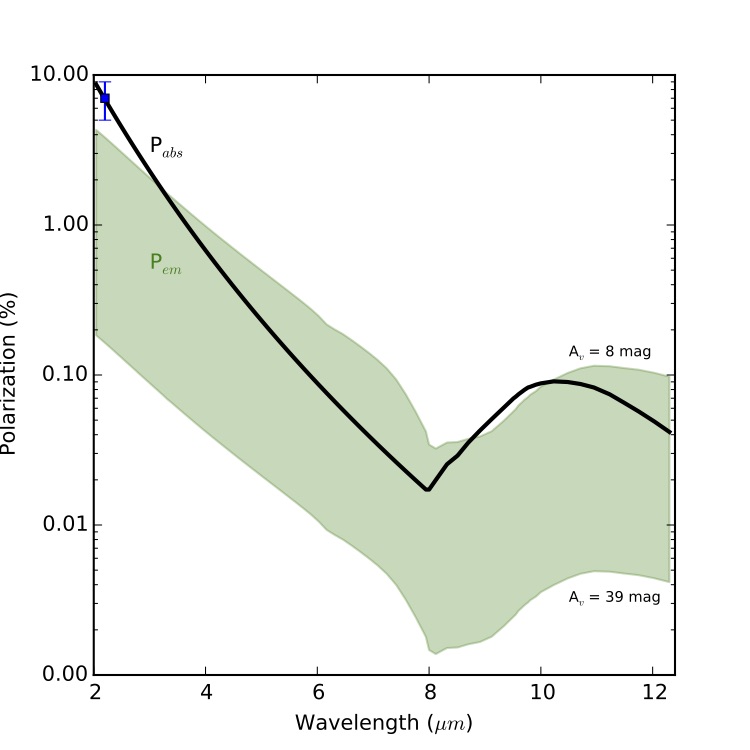}
\caption{Expected absorptive ($P_{\mbox{\tiny abs}}$, black solid line) and emissive ($P_{\mbox{\tiny em}}$, green shaded area) intrinsic polarization from the obscuring dusty structure surrounding the central engine of NGC 1068. The absorptive polarization was scaled to the intrinsic polarization at K$'$ (blue dot) of 7.0 $\pm$ 2.2 per cent estimated by \citet{Lopez-Rodriguez:2015aa}.}
\label{fig6}
\end{figure}

With the constraints on each component given above, we estimated the
absorptive and emissive intrinsic polarization of the obscuring dusty
structure of \ngc\ in the 2$-$13 \um~wavelength range
(Fig. \ref{fig6}). At wavelengths $<$4 \um, the absorptive component
is dominant for all values of A$_{\mbox{v}}$. The emissive component
is dominant for low values of the visual extinction in the wavelength
ranges of $\sim$4$-$8 \um~and $>$11 \um, while both absorptive and
emissive components compete in the $\sim$8$-$10 \um~wavelength
range. Based on this model, the expected polarization from
extrapolating the NIR absorptive component is $\lesssim$0.1 per cent
in the 8$-$13 \um~wavelength range. We note that while there could be
a component of polarized emission from those dust grains, if they are
sufficiently warm to emit in the MIR, the uncertainties in estimating
any contribution are very large. We give a further interpretation of
this feature in Section \ref{ANA_Torus}.


\section{Analysis and discussion}
\label{ANA}

We outline the general physical structure of the central 100 pc of
\ngc, which will aid us in the analysis of the several observed
polarization features introduced in Section \ref{RES}. The polarized
structures are discussed in the following order: 1) the northern
ionization cone, 2) the southern ionization cone, and 3) the obscuring
dusty structure.


\begin{figure*}
\centering
\includegraphics[angle=0,trim=0cm 0.5cm 0cm 1cm,scale=0.23]{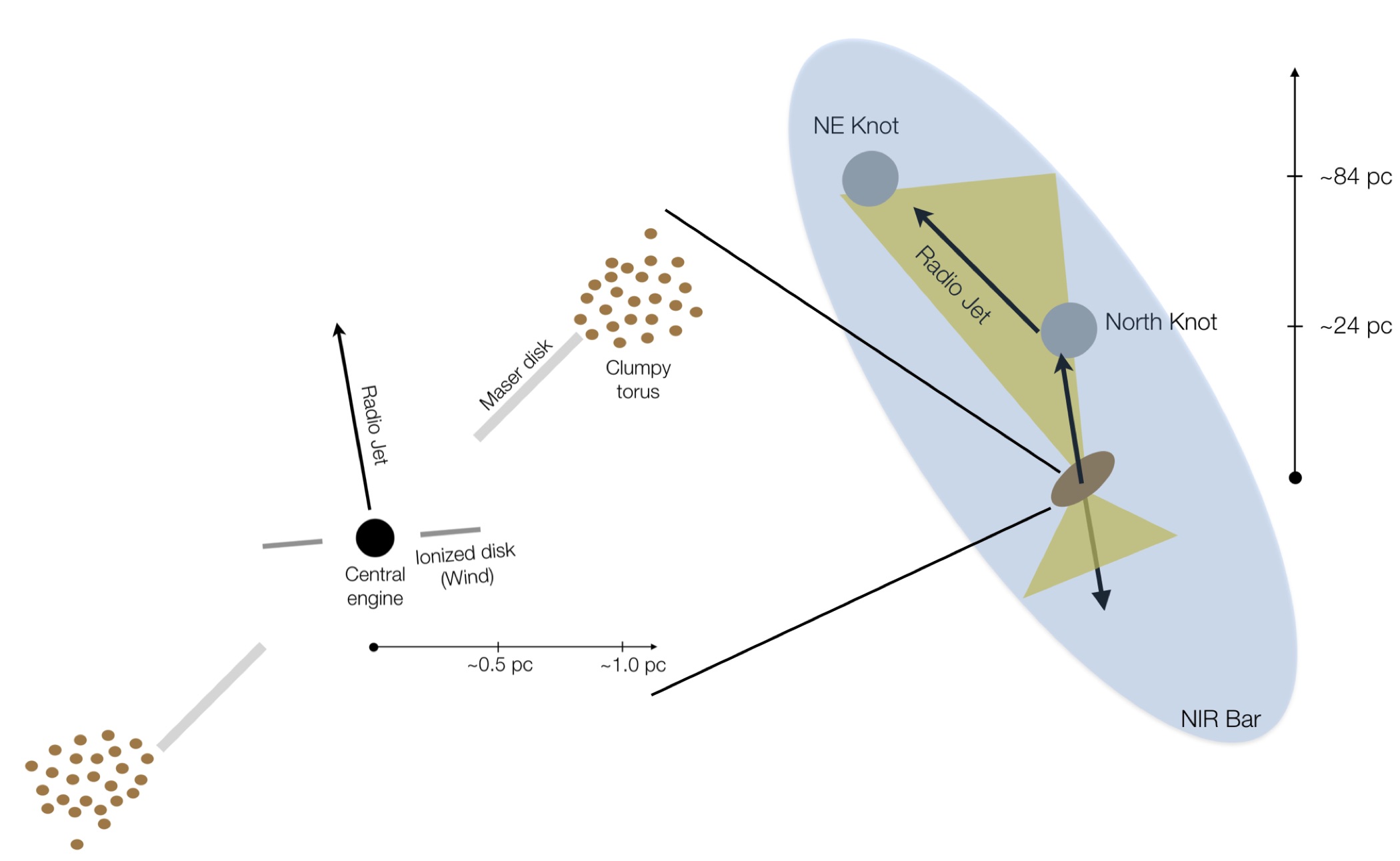}
\caption{Sketch of the central 100 $\times$ 100 pc (right panel) with
  a zoom-in of the central few parsecs (left panel) of \ngc\ (both panels are not on
  linear scale). In the central few parsecs (left
  panel): the central engine (central black dot), the ionized
  disk/wind (dark gray region) occupying the inner $\sim$0.4 pc
  \citep{Gallimore:2004aa}, the maser disk (light gray region) from
  $\sim$0.6 pc to $\sim$1.1 pc at PA =
  $-$40\degr~\citep{Gallimore:2001aa}, the obscuring dusty material
  (brown dots) up to $\sim$1.35 pc at PA =
  $-$42\degr~\citep{Raban:2009aa} and the radio jet (black solid
  arrow) at PA $\sim$ 10\degr\ \citep{Gallimore:2004aa} are shown. In
  the 100 $\times$ 100 pc scale (right panel): the central few parsecs
  are represented as a brown ellipsoid with the PA of the obscuring
  dusty material. The inner radio jet at PA $\sim$ 10\degr\ is shown
  as a black arrow, bending to PA $\sim$ 40\degr\ after interaction
  with the \emph{North knot} (grey circle), towards the NW knot (grey
  circle). The [\ion{O}{III}] ionization cones (yellow polygons) and
  the NIR bar (light-blue ellipse) are shown. North is up and East is
  left.}
\label{fig7}
\end{figure*}


We present the structure of the inner 100 parsecs of
\ngc. It shows a complex geometry in the inner few parsecs. Through 5
and 8.5 GHz radio observations using the very long baseline array
(VLBA), \citet{Gallimore:2004aa} detected the location of the hidden
active nucleus (labeled as S1 in their paper) as a parsec-sized
($\sim$0.8 pc diameter) structure with a major axis at
P.A. $\sim$105\degr. The emission of this component most likely
indicates strong free-free emission from hot (10$^{4}$$-$10$^{5}$
K) ionized gas. This structure is oriented nearly perpendicular to the
inner radio jet axis, P.A.
$\sim$12\degr~\citep{Gallimore:1996aa}.
The kinematics of the maser clouds indicate that the masers are
located in a non-Keplerian disk with inner radius $\sim$0.65
pc and outer radius $\sim$1.1
pc at a P.A. of $-$40\degr~\citep{Gallimore:2001aa}.
Through MIR interferometric observations, \citet{Jaffe:2004aa}
estimated a warm (T $=$
320 K) dust structure with a 2.1 pc thick and 3.4 pc diameter,
surrounding a hot (T $>$
800 K) compact ($\sim$1
pc) structure. Additional MIR interferometric observation by \cite{Raban:2009aa} suggested a hot ($\sim$800
K) 1.35 pc long and 0.45 pc thick structure with a larger 3 $\times$
4 pc and warm (T $\sim$300
K) dust structure at a P.A. = $-$42\degr.
These authors suggested that the dust and the maser disk are
co-spatial, where the dusty torus begins at the outer edge of the
maser disk. These results are consistent with the established
relationship between the water maser excitation and warm ($>$600
K) molecular dust in AGNs \citep{Neufeld:1994aa}. Where the maser disk
ends and the dusty torus structure begins may be more a question of
semantics rather than a true physical boundary. Another implication
from their results is that as the masers are seen edge-on, so is the
torus, consistent with the classification of \ngc\ as a Type~2 AGN. A
sketch of the inner few parsecs of \ngc\ is shown in Figure
\ref{fig7}-left. Note that this figure is a detailed sketch of the
central parsec of \ngc\ adapted from fig.~6 in
\citet{Lawrence:2010aa}. Based on the geometry of the inner few
parsecs, misalignments between these structures are found. Although
misaligned structures on similar spatial scales have been observed in
other AGNs
\citep[i.e.][]{Miyoshi:1995aa,Wilson:1995aa,Greenhill:1997aa,Hagiwara:1997aa,Trotter:1998aa},
it is not the scope of this paper to investigate these, and we refer
to the reader to the discussion in \citet{Lawrence:2010aa}.

At larger scales ($>$10
pc), \ngc\ displays a variety of structures. Most of the radiation
emerges from the interaction of the radio jet with material in the
ionization cones partially extinguished by the disk of the galaxy
and/or the NIR bar. The inner few parsecs of the ionization cone are
roughly aligned with the radio jet in the N-S direction. At
$\sim$24
pc in the northern ionization cone, the radio jet changes direction
when it interacts with the molecular cloud (i.e. \emph{North Knot})
\citep[e.g.][]{Gallimore:2001aa}. The radio-jet emission between the
central few parsecs and the molecular cloud is dominated by
synchrotron emission, while the jet-molecular cloud interaction
radiates at all wavelengths. After this interaction, the radio jet
expands and is roughly aligned with the extended northern ionization
cone on a $\sim$50$-$100
pc scale. The radio jet in the southern ionization cone does not
change direction and moves away from our LOS. The radio jet and molecular cloud are extinguished by the NIR bar
with an extension of $\sim$1.92
kpc at a PA
$\sim$48\degr~\citep{Scoville:1988aa,Schinnerer:2000aa,Emsellem:2006aa}.
Figure~\ref{fig7}-right outlines the large scale structures.



\subsection{The northern ionization cone}
\label{ANA_North}

We analyze here the polarization mechanism from the northern extended
polarization feature, and the \emph{North knot}. As we noted in the introduction, \ngc\ has a NIR bar with an extension of 32 arcsec (1.92 kpc) at a P.A. =
48\degr~\citep{Scoville:1988aa,Schinnerer:2000aa,Emsellem:2006aa}. In
this region, the optical \citep{Scarrott:1991aa} and, J and H
\citep{Packham:1997aa} polarization shows a P.A. of polarization
consistent with the orientation of the NIR bar. This result is most
easily interpreted as polarization arising from extinguished
starlight passing through aligned dust grains in the NIR bar. Dichroic
absorption dominates the polarization in the NIR bar, suggesting that
the dust grains are aligned by the galactic magnetic field at P.A. =
48\degr. The measured P.A. of polarization of $\sim$44\degr~in the
northern ionization cone is consistent with the measured P.A. of
polarization in the optical, J, and H bands. This result suggests that the
dichroic absorption mechanism may dominate from the optical to the MIR
in the NIR bar at the northern ionization cone.

\begin{figure*}
\centering
\includegraphics[angle=0,trim=0cm 1cm 0cm 0cm,scale=0.16]{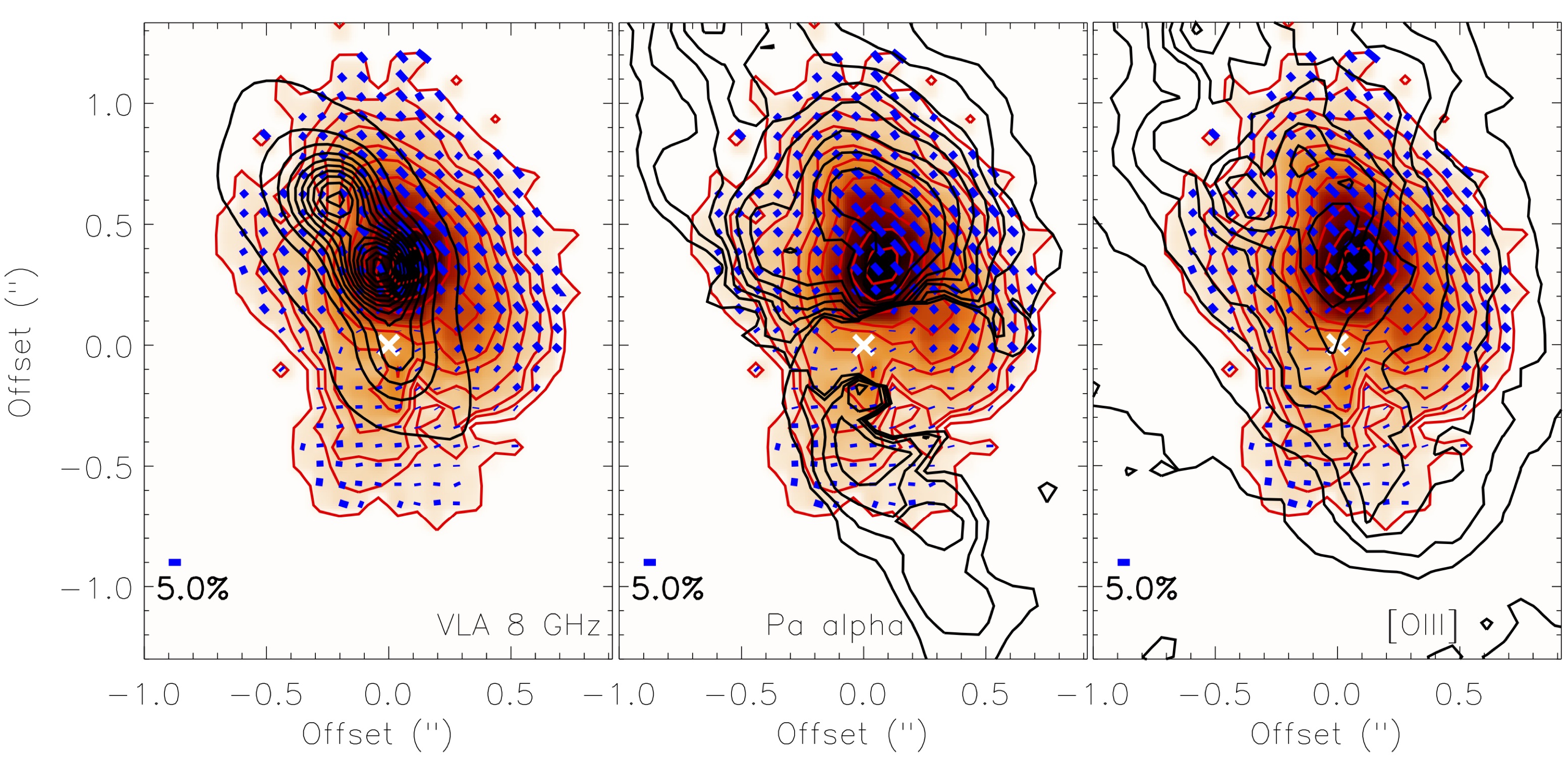}
\caption{The 3 arcsec $\times$ 2 arcsec (180~pc~$\times$~120 pc) central region of NGC 1068 showing the polarized flux image with overlaid polarization vectors at 8.7 \um~(as in Fig. \ref{fig2}) with overlaid contours of VLA 8 GHz (left), Pa${\alpha}$ (middle) and, [\ion{O}{iii}] (right) lines emission. For VLA 8 GHz, contours start at 5$\sigma$ and increase in steps of 50$\sigma$. For Pa${\alpha}$, contours start at $3\sigma$ and increase in steps of $3\sigma$. For [\ion{O}{iii}], contours star at 8$\sigma$ and increase as 2$^{n}\sigma$, with $n = 4, 5, 6, \dots$ North is up and East is left.}
\label{fig8}
\end{figure*}

To understand the polarization of the \emph{North knot} and its relation with
the extended polarization feature, we further compare the degree of
polarization map at 8.7 \um, and maps of the emission lines
Paschen-${\alpha}$ (Pa${\alpha}$) at 1.875
\um~(Fig. \ref{fig8}-middle) and [\ion{O}{iii}] at 0.5007
\um~(Fig. \ref{fig8}-right). We obtained archival images
of \ngc\ at Pa${\alpha}$ (ID: 7215, observed on
1998 Feb. 21) and [\ion{O}{iii}] (ID: 5754, observed on 1995 Jan. 17)
emission lines using \textit{HST}. We found that the MIR polarization
is spatially coincident with the Pa${\alpha}$ (Fig. \ref{fig8}-middle)
and [\ion{O}{iii}] (Fig. \ref{fig8}-right) emission
lines. Pa${\alpha}$ line emission can be interpreted as a tracer of
young ($<$30 Myr) star formation or gas
emission. \citet{Storchi-Bergmann:2012aa} found an episode of young
(30 Myr) stellar population in a ring-like structure at $\sim$100 pc
from the nucleus, coincident with the ring of molecular gas detected
in the 2.12 \um~warm molecular hydrogen line and in cold molecular gas
detected by ALMA \citep{Garcia-Burillo:2014aa}. We found that the
morphology of this young stellar population and cold molecular gas
resembles neither the Pa${\alpha}$ line emission nor the MIR features
in the northern ionization cone. Additionally, we found no evidence of
11.3 \um~PAH feature in the northern ionization cone in our
spectropolarimetric observations (Fig. \ref{fig4}-left), consistent
with MIR spectroscopic observations by \citet{Mason:2006aa}. This comparison between Pa${\alpha}$ line emission with the degree of polarization suggests that the Pa${\alpha}$ line emission is tracing gas
photoionized by the AGN and/or gas excited by shocks in the ionization
cones. This result is further supported by the spatial correspondence
of Pa${\alpha}$ and [\ion{O}{iii}] emission lines, as the latter is a
good tracer of the gas emission in the narrow line regions
\citep[e.g.][]{Evans:1991aa,Emsellem:2006aa,Kraemer:2015aa}. Based on these results, we interpret the MIR polarization of the northern
ionization cone as arising from extinguished radiation from dust and gas heated by the AGN
\citep[e.g.][]{Crenshaw:2000ac,Kraemer:2000ab,Das:2007aa} and the
interaction of the jet \citep[e.g.][]{Gallimore:1996aa} with dust and
gas in the northern ionization cone passing through the aligned dust grains in the
NIR bar producing the uniform P.A. of polarization in the northern
ionization cone. The fact that the P.A. of polarization is not
perfectly aligned along the P.A. of the NIR bar can be explained as
the radiation from the jet-molecular cloud interaction being polarized
with a slightly different P.A. of polarization. Variations of the S/N
in the northern regions can also explain the $\sim$10\degr\
fluctuations of the P.A. of polarization.

Based on the model presented in Section~\ref{PolModel}, the dust
grains in the northern ionization cone are different from those in the
interstellar medium (ISM) in the Milky Way. This confirms the
spectropolarimetric observations of \citet{Aitken:1984aa}, who also
found a relatively featureless polarization spectrum in a 4.2 arcsec
(252 pc) beam. They concluded that it was not produced by standard
silicate grains. The \emph{North knot} region produces the bulk of the
nuclear polarization in \ngc, and our observations confirm that it
mostly arises from non-silicate grains, despite the clear signature of
silicate absorption in the total flux density spectrum
(Fig.~\ref{fig4}, left).

We considered other polarization mechanisms to explain the MIR
polarimetric observations of the northern ionization cone. Scattering
off electrons and/or dust grains can be ruled out due to the uniform
P.A. of polarization in this region; there is no evidence of a
centro-symmetric pattern that would be expected for scattering of
radiation from a nuclear source. We considered also the possibility of
the jet-molecular cloud interaction as a dust grain alignment
mechanism in the northern ionization cone. For this study, we obtained
archival images of \ngc\ at 8 GHz (Observer: AC467, observed on 1999
Sep. 08) using the Very Large Telescope (VLA). We plot the degree of
polarization map at 8.7~\um\ and radio emission at 8~GHz in Figure
\ref{fig6}, left. If the jet-molecular cloud interaction dominated in
the northern ionization cone, we would expect the magnetic field to be
parallel to the interface of the jet-molecular cloud interaction
\citep[e.g.][]{Drury:1983aa}. The magnetic field is compressed in the
interface and increases in strength, resulting in better dust grain
alignment in the northern ionization cone than in the ISM.  As noted
in Section \ref{ANA}, the jet bends at the position of the North
knot, which would imply variations of P.A. of polarization of
$\sim$30\degr~as a consequence of the different interface between jet
and surrounding molecular dust in the northern ionization
cone. Despite the registration uncertainties of $\sim$0.1-0.3 arcsec
in Fig. \ref{fig6}-left, the uniform P.A. extends further
($\sim$1.5~$\times$~1.0~arcsec, i.e. 90~$\times$~60~pc) than the radio
jet emission. These observations suggest that the jet-molecular cloud
interaction cannot be the dominant polarization mechanism in the whole
northern ionization cone.


\subsection{The southern ionization cone}
\label{ANA_South}

\begin{figure}
\centering
\includegraphics[angle=0,trim=1cm 2.cm 0cm 0cm,scale=0.10]{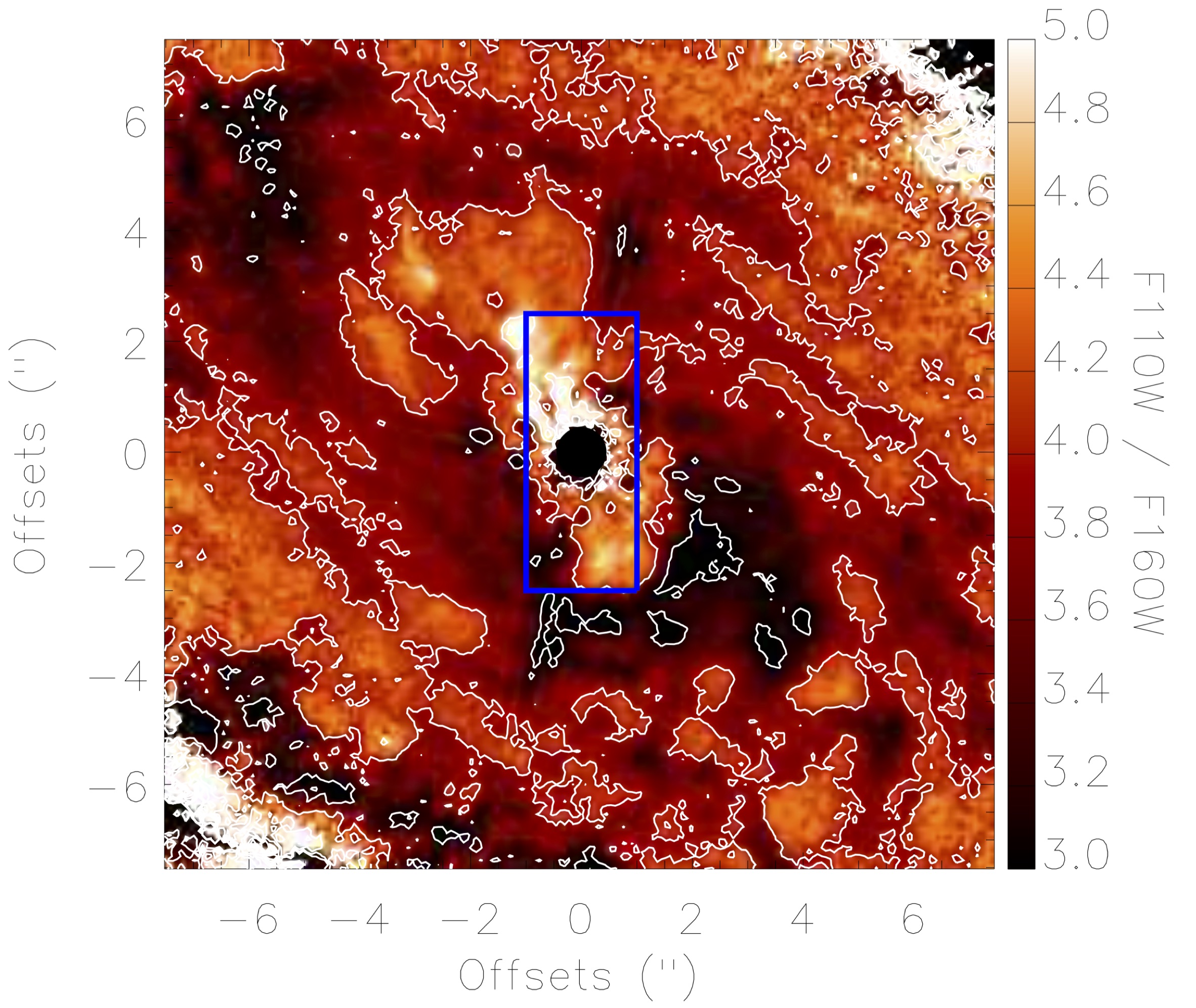}
\caption{F110W / F160W color map of the 15 arcsec $\times$ 15 arcsec (900 pc $\times$ 900 pc) central region of NGC 1068. Contours start at 3$\sigma$ and increase as 1$\sigma$. The NIR images were PSF subtracted using TinyTim, then the colormap was produced. The central 0.4 arcsec (24 pc) was masked (black filled circle) due to the noise pattern produced by the PSF subtraction.The FOV (5 arcsec $\times$ 2 arcsec, 300 pc $\times$ 120 pc) of the MIR observations (blue rectangle) is shown.}
\label{fig9}
\end{figure}

We present here an analysis of the southern ionization cone and its \emph{Southern feature}. The optical and NIR spectral lines and continuum emission are much fainter to the South of the core of \ngc\ \citep[][see also Fig. \ref{fig6}]{Macchetto:1994aa,Bruhweiler:2001aa,Thompson:2001aa},
with a deeper silicate feature, $\tau_{\mbox{\tiny 9.7}}$, within 1
arcsec SSW of the core \citep{Mason:2006aa}. To show these results, we
produced (Fig. \ref{fig9}) a NIR colormap of \ngc\ using archival
\textit{HST}/NICMOS images, specifically F110W and F160W (ID: 7215,
observed on 1998 Feb. 21). Next, we modeled the PSF at each filter
using TinyTim \citep{Krist:2011aa}, and then the PSFs were subtracted
from the \emph{HST}/NICMOS images. Finally, the F110W / F160W
colormap (Fig. \ref{fig9}) was produced. Fig. \ref{fig9} shows larger amount of
host galaxy dust to the South of the core of \ngc, suggesting that the
southern ionization cone suffers from higher extinction than the
northern ionization cone. This result is in agreement with 
the increase of the NIR polarization to the SW
\citep{Young:1995aa,Packham:1997aa,Simpson:2002aa,Lopez-Rodriguez:2015aa},
interpreted as an additional scattering medium in this region. In
addition, the dusty disc of the galaxy has an inclination of
29\degr~\citep*{Garcia-Gomez:2002aa}, screening the southern but not
the northern ionization cone
\citep[e.g.][]{Packham:1997aa,Bock:1998aa,Kraemer:2000aa}. Thus, the
southern ionization cone is more extinguished than the northern one,
as is also suggested by the deeper silicate feature
\citep{Mason:2006aa}.


We found that the MIR polarization of the southern ionization cone is spatially coincident with the
Pa${\alpha}$ (Fig. \ref{fig8}-middle) and [\ion{O}{iii}]
(Fig. \ref{fig8}-right) emission lines. The P.A. of polarization
(Fig. \ref{fig3}-bottom) is $\sim$30\degr~along the Pa${\alpha}$
emission (Fig. \ref{fig8}-middle), whereas a P.A. of polarization of
$\sim$10\degr~is observed where not spatial coincident with this emission line is found. For those regions where MIR polarization and
Pa${\alpha}$ emission are spatially coincident, we interpret the
observed polarization in the southern ionization cone as dust directly
heated by the jet; this dust emission passes through both the NIR bar
and the disc of the galaxy. For those regions where MIR polarization and
Pa${\alpha}$ emission are not spatially coincident, the polarization
arises from dichroic absorption in the galaxy. As the southern
ionization cone is affected by several extinction components, its
P.A. of polarization differs from that observed in the northern
regions.

As in the northern regions, scattering off electrons and/or dust grains can be ruled out, as there is no
evidence of the centro-symmetric pattern that would be expected for
scattering of radiation from a nuclear source. If
the radio jet were responsible for the dust grain alignment, then a
constant P.A. of polarization would be expected. Such a uniform
polarization pattern is however not observed, and thus polarization
arising from dust emission of aligned dust grains by the radio jet can
be ruled out.

\subsubsection{The Southern feature}
\label{ANA_South_feature}

Based on the model presented in Section
\ref{PolModel_Fit1}, we interpret the polarization of the \emph{Southern feature}
as a polarized dust emission component passing through a large
concentration of silicates as well as the dusty disc of the galaxy
along our LOS. The polarized dust emission component in the \emph{Southern feature} can arise from 1) magnetically aligned dust grains
directly heated by the jet in the southern ionization cone close to
the AGN, or 2) aligned dust grains in the obscuring dusty structure
surrounding the central engine (Section \ref{ANA_Torus}). As the core
is included within the aperture used to perform the measurements of
the \emph{Southern feature}, we cannot distinguish between dust grains in the
southern ionization cone and those grains associated with the
circum-nuclear dusty obscuring structure. Unfortunately, to obtain polarimetric measurements in an aperture further south of the nucleus to avoid the core of \ngc\ produced low S/N
spectra making it difficult to interpret the polarization.


\begin{figure*}
\centering
\includegraphics[angle=0,trim=0cm 1cm 0cm 0cm,scale=0.17]{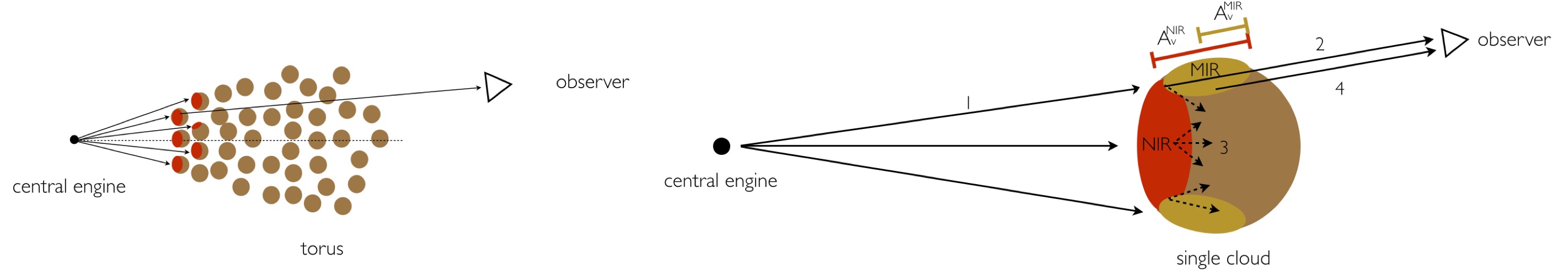}
\caption{Sketch of the clumpy models (left), and clump configuration
  (right) compatible with the observed IR polarization. 1) The
  innermost clump is directly irradiated by the central engine and
  re-emits in the NIR from the directly illuminated face of the
  clump. Then, 2) the NIR radiation passes through the low-density
  regions of the dusty clump, towards the observer. Polarization
  arises from dichroic absorption through a visual extinction
  A$^{\mbox{\tiny NIR}}_{\mbox{\tiny v}}$. 3) The
  radiation going through higher density regions is completely
  extinguished. 4) The shielded outer layer of the clump emits in the
  MIR, and MIR polarization arises from dichroic emission at the low
  extinction regions, A$^{\mbox{\tiny MIR}}_{\mbox{\tiny v}}$, of the
  outer layer of the clump.  }
\label{fig10}
\end{figure*}


As noted in Section \ref{PolModel}, a prominent peak at wavelengths
$\sim$10 \um~would be present
in the degree of polarization if silicate grains were present. It is
interesting that the polarization of the \emph{Southern feature} can be
reproduced with silicate emission/absorption, while the \emph{North knot}
does not show such silicate feature in the polarization spectrum. This
suggests different dust composition within the central 120 pc of
\ngc. If the interaction of the jet with the dust in the northern
ionization cone is changing the dust properties, then the observed
silicate feature of the \emph{Southern feature} may favour the observed
polarization arising from dust grains in the obscuring dusty structure
surrounding the central engine of \ngc. We study the contribution of the obscuring dusty structure in Section
\ref{ANA_Torus}.



\subsection{The obscuring dusty structure}
\label{ANA_Torus}

Although the MIR total flux density peaks at the position of the AGN
\citep[i.e.][]{Braatz:1993aa,Cameron:1993aa,Lumsden:1999aa,Bock:1998aa,Alloin:2000aa,Bock:2000aa,Tomono:2001aa,Jaffe:2004aa,Galliano:2005aa,Mason:2006aa,Tomono:2006aa,Raban:2009aa,Lopez-Gonzaga:2014aa},
a very low polarization with an upper-limit of 0.3 per cent at this
location is detected within the 8$-$13 \um~wavelength range
(Fig. \ref{fig2} and \ref{fig3}). We present here an interpretation of the MIR polarization of the
obscuring dusty structure of \ngc\ in the context of the
\C\footnote{\C\ models can be found at: \url{www.clumpy.org}} torus
models \citep{Nenkova:2002aa,Nenkova:2008ab,Nenkova:2008aa}. \C\ models the obscuring dusty structure as an optically and geometrically thick torus surrounding the central engine of an AGN,
and comprised of dusty clouds.

\citet{Lopez-Rodriguez:2015aa} interpreted the NIR polarization of
\ngc\ in terms of the level of magnetization through the estimation of
the thermal-to-magnetic pressure ratio in the clumps. Specifically, the NIR polarization they measured is
determined in the outermost material of lower optical depth in the
clumps. Based on these results, and the fact that the
absorptive/emissive polarization components should arise from the same
aligned dust grains, but with orthogonal polarization angles, we are
able to give an interpretation of the IR polarization. The NIR
polarization arises from the passage of NIR radiation from directly
illuminated clumps through the low-density dusty regions of the clumps
with a visual extinction of A$^{\mbox{\tiny NIR}}_{\mbox{\tiny v}}$.
The same aligned dust grains emit at MIR wavelengths, and this
radiation passes through lower extinction,
A$^{\mbox{\tiny MIR}}_{\mbox{\tiny v}}$, than the NIR photons. The
lower the extinction, the higher the emissive intrinsic polarization
component, i.e.  P$_{\mbox{\tiny em}} = $
P$_{\mbox{\tiny abs}} / \tau$. Since our observations do not show a
MIR polarization at the core, and from our polarization model, the
minimum visual extinction that the MIR radiation passes through is
A$^{\mbox{\tiny MIR}}_{\mbox{\tiny v}} = $ 8 mag. For lower
A$^{\mbox{\tiny MIR}}_{\mbox{\tiny v}}$, a degree of polarization
higher than 0.1 per cent is expected (Fig. \ref{fig6}), which would be
detectable in our observations. Thus, the model of emissive
polarization arising from aligned dust grains in the clumps is
consistent with the non-detection of MIR polarization at the core of
\ngc. Based on these results, Figure~\ref{fig10} shows the possible
configuration for a clump compatible with the observations.

Although the \C\ models assume only discrete dust clumps with no
inter-clump medium, an inter-clump medium should be considered as a more physical realistic model. We here put constraints on the extinction of the inter-clump medium to be used in two-phase clumpy torus models. Assuming a magnetohydrodynamical framework in the obscuring dusty structure
of \ngc\ \citep{Lopez-Rodriguez:2015aa}, a global magnetic field at
the same location of the obscuring dusty material aligns the dust
grains in both the inter-clump medium and in the clumps. In this
scheme, the IR polarization can arise from the passage of radiation
through aligned dust grains in the clumps, as described above, and/or
in the the inter-clump medium. To obtain an upper-limit on the extinction of the inter-clump medium, we assume that the clumps are optically thick with uniform density, then the aligned dust grains in
the inter-clump medium would be responsible for the observed dichroic
polarization at IR. Specifically, the IR radiation from the clumps passes through the aligned dust
grains in the inter-clump medium. We now define the ratio
of the optical depth of the clump, $\tau^{\mbox{\tiny c}}$, and
inter-clump medium, $\tau^{\mbox{\tiny m}}$, with the contrast
parameter C = $\tau^{\mbox{\tiny c}}$/$\tau^{\mbox{\tiny m}}$. C is a
positive quantity, with C = 1, describing the obscuring dusty material
in a smooth distribution of dust, where large values of C will enhance
the clumpy distribution of dust grains. An optical depth of the clumps of $\tau_{\mbox{\tiny v}}^{\mbox{\tiny c}} =$ 49$^{+4}_{-3}$ was took from \citet{Alonso-Herrero:2011aa} using \C\ models to fit the
nuclear IR spectral energy distribution (SED) of \ngc. Based on
our polarization model, the minimum visual extinction to be compatible with
the observations is A$_{\mbox{\tiny v}} =$ 8 mag, or
$\tau_{\mbox{\tiny v}} =$ 8.7. Taking this value as the extinction of
the inter-clump medium, we obtain a contrast of C =
5.6$^{+0.4}_{-0.7}$ at V-band. This result favors the clumpy distribution of
dust versus a smooth-density obscuring material. The clumpy models \citep{Stalevski:2012aa,Siebenmorgen:2015aa} including a two-phase
medium are crucial to generate SED based on
the MIR polarimetric observational constraints. However, the
development of these models is beyond the scope of this paper.



\section{Conclusions}
\label{CON}

We presented sub-arcsecond resolution MIR (7.5$-$13 \um) imaging- and
spectro-polarimetric observations of \ngc\ obtained with
\hbox{CanariCam} on the 10.4-m GTC. Using the wide MIR wavelength
coverage, we were able to disentangle several polarized structures in
the complex 3~$\times$~2~arcsec (180~$\times$~120~pc) central region
through their polarization signatures. At all wavelengths, we found:

\begin{enumerate}

\item Northern ionization cone: the most prominent feature is the
  uniform P.A. of polarization, $\sim$44\degr, with an extension of
  $\sim$1.5~$\times$~1.0~arcsec (90~$\times$~60~pc). Based on our
  polarization model, we found that the polarization arises from dust
  and gas emission in the ionization cone, heated by the AGN and jet,
  and further extinguished by aligned dust grains in the NIR bar. In
  this region, we found that the \emph{North knot}, located at
  $\sim$0.4 arcsec (24 pc) from the core, is spatially coincident with
  the interaction of the jet and molecular cloud. The \emph{North knot} shows
  a uniform P.A. of polarization and a slow decrease in the degree of
  polarization with increasing wavelength. No silicate feature is
  found in the polarization spectrum of the \emph{North knot}, suggesting
  that the dust grains are different from those in the ISM.

\item Southern ionization cone: we found a polarized feature at
  $\sim$0.16 arcsec (9.6 pc) South of the core. Based on our
  polarization model, the \emph{Southern feature} polarization arises from an
  emissive polarization component passing though a larger
  concentration of dust than in the northern ionization cone. For this
  emissive component, we cannot distinguish between 1) magnetically
  aligned dust grains directly heated by the jet in the southern
  ionization cone close to the AGN, and 2) magnetically aligned dust
  grains in the obscuring dusty structure surrounding the central
  engine. The polarization spectrum shows a prominent silicate feature, 
  suggesting that different dust grains composition or sizes may be present in the northern and
  southern ionization cones.

\item Obscuring dusty structure: An upper limit of 0.3 per cent in the
  degree of polarization from the core is detected in our
  observations. Based on our polarization model, the expected
  polarization from extrapolating the NIR absorptive component is
  $\lesssim$0.1 per cent in the 8$-$13 \um~wavelength range. We note
  that while there could be a component of polarized emission from
  those dust grains, if they are sufficiently warm to emit in the MIR,
  the uncertainties in estimating any contribution are very large. We
  suggested that the MIR polarization arises from the passage of MIR
  radiation through aligned dust grains in the shielded edges of the
  clumps located at the innermost regions of the obscuring dusty
  structure (Fig. \ref{fig10}).
  Alternatively, an obscuring clumpy structure with an inter-clump
  medium, and with an optical depth contrast between the clumps and
  the inter-clump medium of C = 5.6$^{+0.4}_{-0.7}$ at V-band, can also 
  reproduce the upper-limit of the degree of polarization in our observations.
\end{enumerate}

The sub-arcsecond resolution polarization observations of \ngc\ demonstrate that there are a number of different regions and mechanism operating within the central 10s of parsecs in this AGN. Similar complexity may occur in other AGNs, where the current observations cannot resolve these structures. Further sub-arcsecond MIR imaging- and spectro-polarimetric observations to several AGNs are crucial to find ordinary and/or extraordinary properties in AGN allowing us to refine and/or modify the AGN unified model.


\section*{Acknowledgements}

It is a pleasure to acknowledge discussion with R. Mason. We would
like to thank the anonymous referee for their useful comments, which
improved the paper significantly. Based on observations made with the
Gran Telescopio CANARIAS (GTC), installed in the Spanish Observatorio
del Roque de los Muchachos of the Instituto de Astrof\'isica de
Canarias, in the island of La Palma. E.L.R and C.P. acknowledge
support from the University of Texas at San Antonio. C.P. acknowledges
support from NSF-0904421 grant. A.A.-H. acknowledges financial support
from the Spanish Ministry of Economy and Competitiveness (MINECO)
under the 2011 Severo Ochoa Program MINECO SEV-2011-
0187. A.A.-H. acknowledges financial support from the Spanish Ministry
of Economy and Competitiveness through grant AYA2012-31447, which is
party funded by the FEDER program, P.E. from grant AYA2012-31277, and
L.C. from grant AYA2012-32295. R.N. acknowledges support by FONDECYT
grant No. 3140436. C.R.A. is supported by a Marie Curie Intra European
Fellowship within the 7th European Community Framework Programme
(PIEF-GA-2012-327934). E.P. acknowledges support from NSF grant
AST-0904896.



\bibliographystyle{mnras}
\bibliography{LopezRodriguez_References_R2}


\bsp	
\label{lastpage}
\end{document}